\documentclass[
 reprint,
 superscriptaddress,
 groupedaddressd,
 nofootinbib,
 amsmath,
 amssymb,
 amsthm
 aps,
 prx,
 floatfix
 longbibliography
]{revtex4-2}

\usepackage{graphicx}
\usepackage{acronym}
\usepackage{dcolumn}
\usepackage{bm}
\usepackage{hyperref}
\usepackage{xcolor}
\usepackage{xspace}
\hypersetup{
    colorlinks,
    linkcolor={red!50!black},
    citecolor={blue!50!black},
    urlcolor={blue!80!black}
}
\usepackage[capitalize]{cleveref}
\usepackage[mathlines]{lineno}

\newcommand{\changed}[1]{\textcolor{black}{#1}}

\newcommand{\pycbc}{\texttt{PyCBC}\xspace}
\newcommand{\gstlal}{\texttt{GstLAL}\xspace}
\newcommand{\spiir}{\texttt{SPIIR}\xspace}
\newcommand{\mbta}{\texttt{MBTA}\xspace}
\newcommand{\cwb}{\texttt{Coherent WaveBurst}\xspace}

\AtBeginDocument{}

\newcommand{\ds}{\ensuremath{\rho}\xspace}
\renewcommand{\Pr}{\ensuremath{\textrm{Pr}}\xspace}
\newcommand{\FAR}{\ensuremath{\textrm{FAR}}\xspace}
\newcommand{\pastro}{\ensuremath{p_{\rm astro}}\xspace}
\newcommand{\pbbh}{\ensuremath{p_{\rm BBH}}\xspace}
\newcommand{\pnsbh}{\ensuremath{p_{\rm NSBH}}\xspace}
\newcommand{\pbns}{\ensuremath{p_{\rm BNS}}\xspace}

\newtheorem{definition}{Definition}

\newacro{CP}{Conformal Prediction}
\newacro{MCP}{Mondrian Conformal Prediction}
\newacro{CBC}{Compact Binary Coalescence}
\newacro{FAR}{False Alarm Rate}
\newacro{iFAR}{inverse False Alarm Rate}
\newacro{MDC}{Mock Data Challenge}
\newacro{ROC}{Receiver Operator Curve}
\newacro{SNR}{Signal-To-Noise Ratio}

\begin{document}

\title{Calibrating gravitational-wave search algorithms with conformal prediction}

\author{Gregory Ashton}
\email{gregory.ashton@rhul.ac.uk}
\affiliation{Department of Physics, Royal Holloway, Egham, TW20 0EX}

\author{Nicolo Colombo}
\affiliation{Department of Computer Science, Royal Holloway, Egham, TW20 0EX}

\author{Ian Harry}
\affiliation{University of Portsmouth, Portsmouth, PO1 3FX, United Kingdom}

\author{Surabhi Sachdev}
\affiliation{School of Physics, Georgia Institute of Technology, Atlanta, GW 30332, USA}

\date{\today}

\begin{abstract}
In astronomy, we frequently face the decision problem: does this data contain a signal? Typically, a statistical approach is used, which requires a threshold. The choice of threshold presents a common challenge in settings where signals and noise must be delineated, but their distributions overlap. Gravitational-wave astronomy, which has gone from the first discovery to catalogues of hundreds of events in less than a decade, presents a fascinating case study. For signals from colliding compact objects, the field has evolved from a frequentist to a Bayesian methodology. However, the issue of choosing a threshold and validating noise contamination in a catalogue persists. Confusion and debate often arise due to the misapplication of statistical concepts, the complicated nature of the detection statistics, and the inclusion of astrophysical background models. We introduce Conformal Prediction (CP), a framework developed in Machine Learning to provide distribution-free uncertainty quantification to point predictors. We show that CP can be viewed as an extension of the traditional statistical frameworks whereby thresholds are calibrated such that the uncertainty intervals are statistically rigorous and the error rate can be validated. Moreover, we discuss how CP offers a framework to optimally build a meta-pipeline combining the outputs from multiple independent searches. We introduce CP with a toy cosmic-ray detector, which captures the salient features of most astrophysical search problems and allows us to demonstrate the features of CP in a simple context. We then apply the approach to a recent gravitational-wave Mock Data Challenge using multiple search algorithms for compact binary coalescence signals in interferometric gravitational-wave data. Finally, we conclude with a discussion on the future potential of the method for gravitational-wave astronomy.
\end{abstract}

\maketitle

\section{Introduction}

The burgeoning field of gravitational-wave astronomy is in a state of rapid evolution.
Second-generation detectors \citep{LIGO, Virgo, KAGRA} have progressed from the first observation of a binary black hole merger \citep{GW150914} to the compilation of extensive transient event catalogues \citep{GWTC3, Nitz:2021zwj, Olsen:2022pin} including also binary neutron star and black-hole neutron-star mergers. With this progress, the methodologies for evaluating the statistical significance of \ac{CBC} signals have undergone notable transformations. While the significance of the initial detection \citep{GW150914} was assessed through the frequentist \ac{FAR}, contemporary catalogues \citep{GWTC3, Nitz:2021zwj, Olsen:2022pin} now use probabilistic Bayesian methods.

However, astrophysicists aiming to learn from gravitational-wave data are confronted with a challenge: the difficulty in identifying signals when their distribution and the noise distributions overlap.
This issue is by no means unique in astronomy (see, e.g. \citet{feigelson2012statistical}). However, gravitational-wave astronomy is an especially intriguing case study because the \ac{SNR} ratio of sources is low, but the potential scientific reward is high.
Moreover, much of the insights derive from studying the population of identified sources \citep{KAGRA:2021duu}.
The events producing signals within current sensitivities are isotropically distributed, so the number of detections scales with the cube of the horizon distance (a measure of the detector sensitivity).
Therefore, there are always more events just beyond the horizon than within: increasing the horizon distance by just 25\% will double the number of events.
The conundrum facing anyone wishing to utilise the hundreds of sources now reported is how to select a threshold to cut between the signals and the noise.
On the one hand, we can choose a conservative threshold, ensuring a high catalogue purity (the fraction of true signals).
However, the conservative threshold also entails a loss of accuracy; after all, we must discard many low-significance astrophysical signals buried in the noise.
On the other hand, choosing a liberal threshold would include a larger number of astrophysical signals but at the cost of bias induced by non-astrophysical catalogue contamination.

Along with the threshold problem, difficulties arise from concurrently applying multiple search algorithms (hereafter referred to as \emph{pipelines}).
The GWTC catalogues produced by the LIGO Scientific, Virgo, and KAGRA (LVK) collaborations (e.g. \citet{GWTC3}) include results from several independent pipelines (specifically, \gstlal \citep{Messick:2016aqy, Sachdev:2019vvd, Tsukada:2023edh, Cannon:2020qnf, Ewing:2023qqe, Sakon:2022ibh}, \mbta \citep{Adams:2015ulm, Aubin:2020goo}, \pycbc \citep{Allen:2004gu, Canton:2014ena, Usman:2015kfa, nitz_2017, Davies:2020tsx}, \spiir \citep{Luan:2011qx, Chu:2020pjv}, and \cwb \citep{Klimenko:2011hz}).
For a given candidate event, the significance between pipelines can vary substantially, reflecting inherent uncertainty in the significance estimate and varying pipeline performance.
However, for those not intimately knowledgeable about the ever-evolving internal workings of the pipelines, it is hard to know when a particular pipeline is more reliable or more sensitive than another.

There are efforts underway to address these issues. For example, population-level analyses can utilise hierarchical models to assess mixed catalogues of signals and noise, avoiding the contamination problem altogether \citep{Gaebel:2018poe, Galaudage:2019jdx, Roulet:2020wyq, Heinzel:2023vkq} and recent efforts are also underway to produce a unified significance estimate \citep{Banagiri:2023ztt}.
Nevertheless, the problem of choosing thresholds will continue to be of interest as mixed methods are in their infancy, and some of the most interesting events will inevitably come from close to the detection horizon: the question of ``does this data contain an astrophysical signal?'' will inevitably persist.

This work will introduce a new and transformative framework to solve this problem using \ac{CP} \citep{vovk2005algorithmic}.
CP is an approach to uncertainty quantification developed within the context of Machine Learning (ML).
CP takes an existing point-prediction algorithm and a calibration data set (consisting of correctly labelled data) and generalises the underlying algorithm's point-prediction to a \emph{prediction set} with a guaranteed \emph{validity} (where valid means that the true label is guaranteed to belong to the set with a predefined confidence).
Its appeal arises from its universal applicability, guarantees, and single assumption: exchangeability of the data.
Moreover, the prediction guarantees are distribution-free: there is no asymptotic assumption or underlying model.
It can be used for classification and regression or, correspondingly, search/detection and inference/measurement in the language of gravitational wave astronomy \citep{Finn:1992wt}.
This work will explore the classification (or search/detection) problem.
We will demonstrate how \ac{CP} can be applied to calibrate pipelines without requiring knowledge of its internal behaviour.
Moreover, we will discuss how \ac{CP} offers an alternative approach to developing a meta-pipeline: taking the inputs from multiple search algorithms and providing a single statement which optimally combines their outputs and is well calibrated.

As we will show, \ac{CP} is simple to implement, easily tested, has minimal assumptions, and no required astrophysical model.
For these reasons, we anticipate that \ac{CP} will be of general interest to the field.
While we will discuss \ac{CP} exclusively in the context of searching for \ac{CBC} signals, we anticipate it will find utility for searches for other sources of gravitational-wave radiation and beyond.

The remainder of this article is structured as follows. In \cref{sec: quant}, we introduce the existing traditional approaches for significance estimation within gravitational-wave astronomy and further motivate this work by considering their real-world performance.
We provide a lay guide to \ac{CP} in \cref{sec: cp}. We apply it in \cref{sec: cpapp} to a toy cosmic-ray detector problem to demonstrate the basic algorithm and extensions in the noise-dominated regime.
Moreover, we also use our toy problem to explain some of the subtleties of \ac{CP}.
In \cref{sec: cbc}, we then go on to apply \ac{CP} to the recent Mock Data Challenge of LIGO-Virgo data \citep{Chaudhary:2023vec}.
Finally, we end with a discussion on the advantages, difficulties, and future prospects of \ac{CP} for gravitational-wave astronomy in \cref{sec: discussion}.

\section{Methodology: Quantifying significance with traditional approaches}
\label{sec: quant}
To begin our discussion, we first review the data, search algorithms, detection statistics, and two dominant quantities used to assess candidate significance: \ac{FAR} and \pastro.
Gravitational-wave strain data comprises quasi-stationary coloured-Gaussian background noise, astrophysical signals, and a variety of non-astrophysical transient noise sources termed \emph{glitches} \citep{Davis:2022dnd, LIGOScientific:2019hgc}.
Absent glitches, the optimal detection statistic is the coloured Gaussian noise matched-filter \ac{SNR}.
When the signal source properties (e.g., the mass of the system) are unknown (as is typical), a bank of templates is searched, often in combination with techniques to maximise or marginalise over subsets of the full parameter space (see, e.g. \citet{Sathyaprakash:1991mt}).
However, in the presence of glitches, the optimal statistic is unknown.
To guide the reader on how the leading searches remain sensitive to astrophysical signals despite frequent glitches, we now describe in broad terms a typical search algorithm or pipeline: the interested reader may wish to review \citet{LIGOScientific:2019hgc} for a deeper discussion.

\changed{The central tools used by most pipelines to distinguish between signals and glitches are the coincidence between detectors and signal consistency checks such as the $\chi^2$ detection statistic \citep{Allen:2004gu}}, which discriminates cases where the data is likely to contain a glitch by analysing the way power is distributed in the broadband signal.
Typically,  the $\chi^2$ and matched-filter \ac{SNR} are combined to produce a \emph{combined ranking statistic} which we label $\ds$.
Additional terms may also be included in the combined ranking statistics, such as weights based on whether the region of parameter space is expected to contain more astrophysical signals and amplitude-phase-time consistency checks between detectors.
The combined ranking statistic can be tuned to maximise the separation of signals from noise (as verified by simulations).
Since the combined ranking statistic is ad-hoc, its background distribution (where the background is taken to mean in the absence of any astrophysical signal) is inherently unknown and must be empirically estimated from the data.
However, gravitational-wave detectors cannot be shielded from astrophysical signals. 
Therefore, pipelines use approaches such as \emph{time-sliding} between separate independent detectors to destroy correlations between astrophysical signals (see, e.g. \citep{LIGOScientific:2007npa, Was:2009vh}), resulting in empirical measurement of the background.
We denote such a background as the set $\left\{\ds\right\} = \left\{\ds_0, \ds_1, \ldots, \ds_{n-1} \right\}$ of $n$ values measured on the background.

Once the background has been estimated for a new candidate event with ranking statistic $\ds'$, the pipeline estimates its significance by calculating the FAR.
Informally, the \ac{FAR} is the amount of background data one must observe to see a ranking statistic as large as $\ds'$.
Such a dimensionful approach results in an intuitive understanding of the significance given knowledge of the amount of data searched.
E.g. for a search of one month of data, an event with a \ac{FAR} of 1 per millennia is a clear detection, while a \ac{FAR} of 1 per day is more likely to be noise.
More precisely, the \ac{FAR} is calculated empirically as the inverse of the number of background events with a ranking statistic of $\ds'$ or greater divided by the segment duration used in the search.
One sees then that the \ac{FAR} is the one-sided right-tail empirical $p$-value divided by the segment duration:
\begin{equation}
    \FAR = \frac{1}{T}\Pr(\ds > \ds' | H_0) = \frac{1}{T}\frac{| \left\{\ds_i : \ds_i > \ds'\right\} |}{|\{\ds\}|}\,,
    \label{eqn: FAR}
\end{equation}
where $H_0$ is the null hypothesis, we apply set-builder notation, and define the set size by $| \cdot |$.

The \ac{FAR} of the first detection was reported in the paper abstract: ``less than 1 event per 203 000 years'' \citep{GW150914}.
However, once a population of signals was established, it became preferential to move to a probabilistic approach instead.
Following \citet{Farr:2013yna}, the foreground and background distributions are modelled by a Poisson mixture model with prior choices informed by the pipeline outputs and previously observed signals.
From this, each pipeline produces a new significance estimate, \pastro: the probability that the signal is astrophysical \citep{Kapadia:2019uut, Dent:2021aaa, Andres:2021vew, Ray:2023nhx}.
Moreover, the modelled approach allows further sub-classification as $\pastro = \pbns + \pnsbh + \pbbh$ (and a complementary probability of terrestrial origin).
With this new approach, the first Gravitational Wave Transient Catalogue (GWTC-1) \citep{GWTC1} defined ``GW'' events as those with a \ac{FAR} less than 1 per 30 days and a \pastro greater than $1/2$. This latter definition has become a de facto standard. For example, a \pastro greater than $1/2$ is the threshold used to identify events for further follow-up in several recent catalogues \citep{GWTC3, Nitz:2021zwj, Olsen:2022pin}).
Yet, it demonstrates that even with a probabilistic interpretation of the nature of a candidate, researchers still like to establish a threshold and draw a clear delineation, and it is quite common to see astrophysics research take the provided thresholds at face value.

The final complicating piece of this picture is that multiple pipelines analyse the same data. Our typical pipeline above described the core features, but each employs a unique arsenal of techniques built over many years by many people.
The result is that for any given candidate, we end up with multiple estimates of its significance: a \ac{FAR} and \pastro per-pipeline.
The pipelines broadly agree for unambiguous signals and noise events where apples-to-apple comparisons can be made.
However, it is in the grey middle ground where things become complicated.
To demonstrate this, we use data from the recent GWTC-3 catalogue \citep{GWTC3}, which reported on data from the second part of the third LIGO-Virgo observing run.
We use the associated data release, which includes triggers where at least one pipeline had a \ac{FAR} of less than 2 per day: as such, we expect this to include both the astrophysical signals and a great number of non-astrophysical candidates.
\begin{figure}
    \centering
    \includegraphics[width=8.6cm]{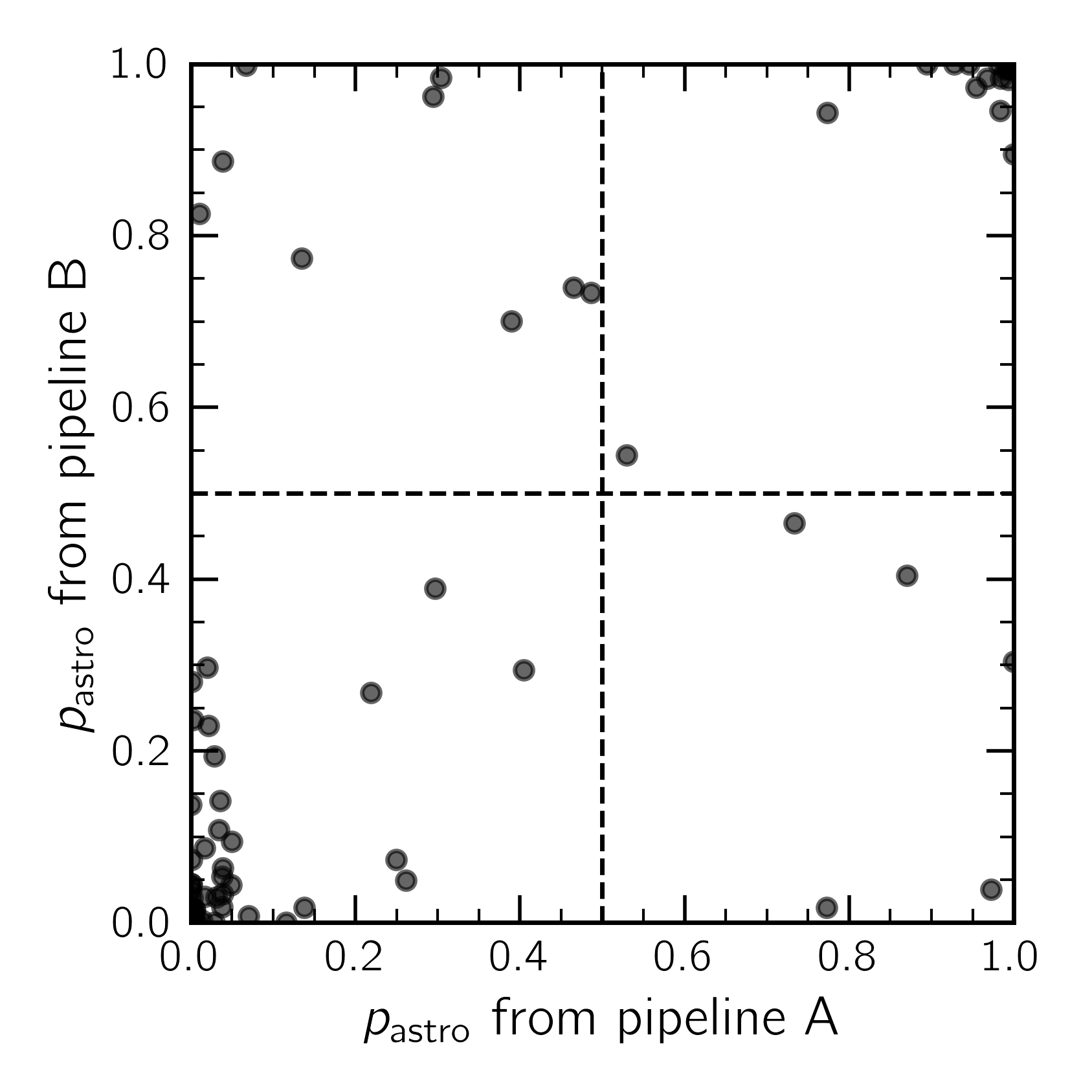}
    \caption{Comparison of the probability of astrophysical origin estimated by pairs of pipelines for all candidates reported in GWTC-3 (including sub-threshold candidates). While clear signal (top right) and clear noise (bottom left) cases usually agree, a significant off-diagonal scatter remains between these points.}
    \label{fig: pastro-pastro}
\end{figure}

In \cref{fig: pastro-pastro}, we scatter-plot the \pastro of each trigger for pairs of \ac{CBC} search pipelines used in GWTC-3 (we exclude the \cwb pipeline that applies an unmodelled search approach).
In the off-diagonal corners, two dense regions correspond to the clear signal (top right) and clear noise (bottom left) cases where pipelines agree.
However, scattered through the plane are confusion cases where one pipeline finds $\pastro > 0.5$, \changed{indicating the data contained an astrophysical source, while the other pipeline is more pessimistic ($\pastro < 0.5$)}.
If we are lucky enough to know experts from both pipelines, we can understand the cause of the discrepancy.
Sometimes, it is well understood different choices lead to different sensitivities in different parts of the parameter space.
If the more sensitive pipeline found the event while the other did not, this explains the difference, and we may gain confidence that this is an astrophysical signal.
Other times, the differences are more contentious or yet to be understood --- this should be expected, as these are complicated multi-stage pipelines with differing and often implicit assumptions. 
Nevertheless, it leaves the uninformed with the previously described choice-of-threshold conundrum exacerbated by the need to learn the detailed inner workings of the pipeline to understand the results.
One standard solution is to take the maximum \pastro, implicitly trusting that the only explanation is variations in sensitivity.
However, another explanation is random uncertainty in significance or even that one pipeline is malperforming.

\begin{figure}
    \centering
    \includegraphics[width=8.6cm]{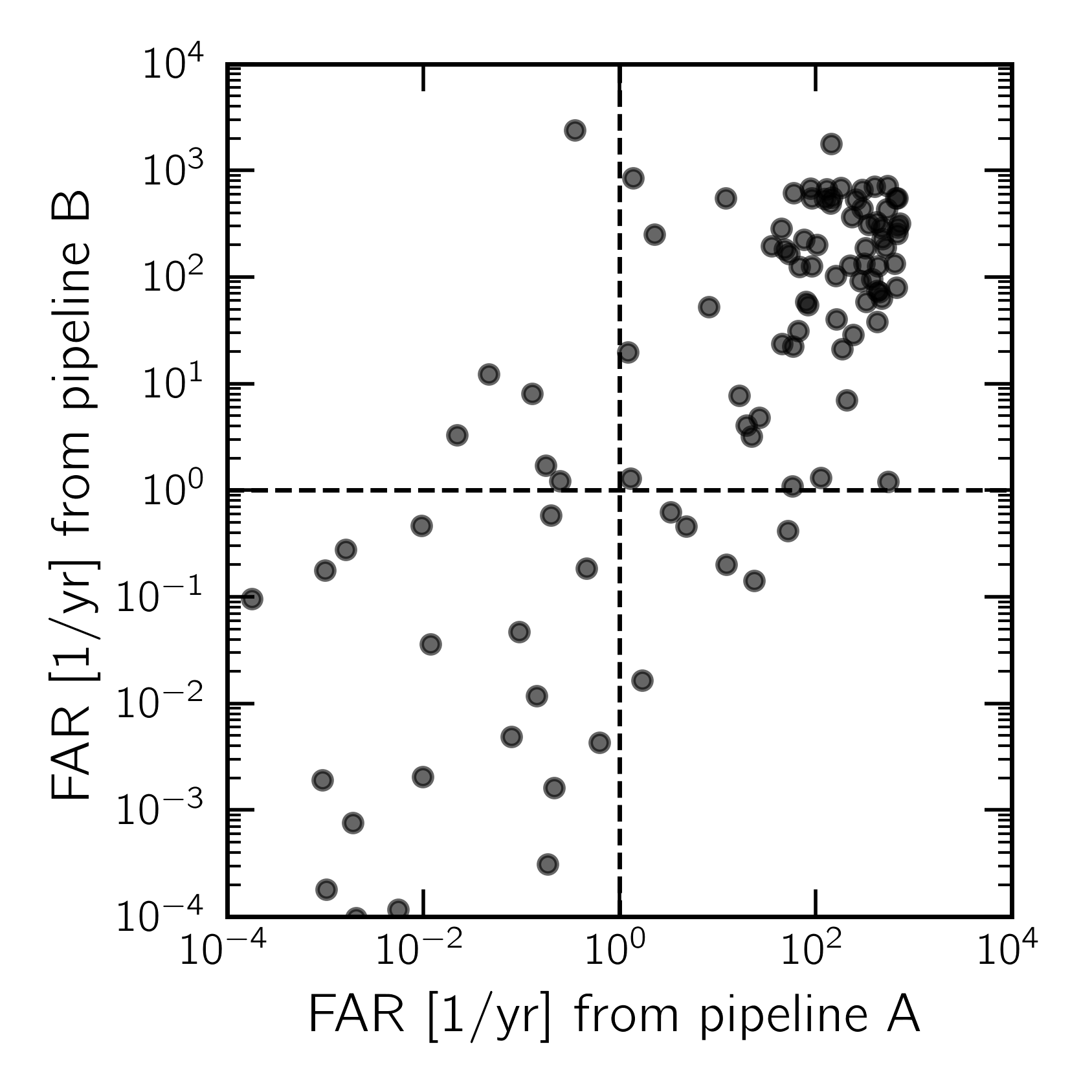}
    \caption{Comparison of the \ac{FAR} estimated by pairs of pipelines demonstrating the intrinsic scatter for all candidates reported in GWTC-3 (including sub-threshold candidates). While clear signal (bottom left) and clear noise (top right) cases usually agree, a significant off-diagonal scatter remains between these points.}
    \label{fig: far-far}
\end{figure}

\begin{figure}
    \centering
    \includegraphics[width=8.6cm]{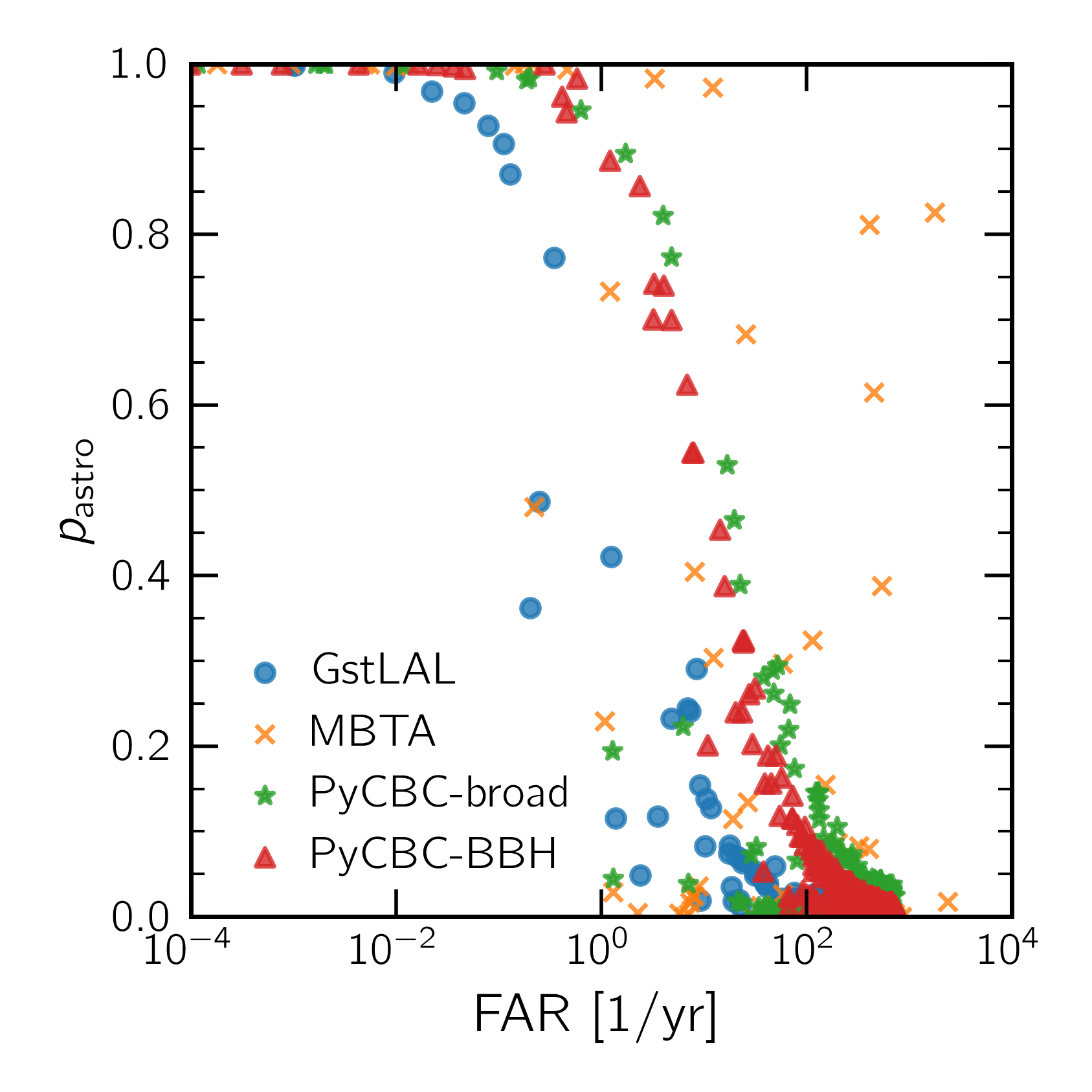}
    \caption{Comparison by pipeline between \pastro and \ac{FAR} for all candidates reported in GWTC-3 (including sub-threshold candidates).}
    \label{fig: FAR-pastro}
\end{figure}

One may imagine that the inclusion of different astrophysical foreground prior models in the Bayesian analysis may explain the scatter in \cref{fig: pastro-pastro} between pipelines; however, \cref{fig: far-far} demonstrates that the scatter is also inherent in the underlying and simpler FAR.
Finally, in \cref{fig: FAR-pastro}, we plot each pipeline's \ac{FAR} against \pastro. Here, we see the approximate sigmoid relationship with significant scatter.

The GWTC-3 results demonstrate the inherent difficulty facing anyone wishing to select a set of events for further analysis.
However, these results are only part of the picture. They present only the pipelines used by the LVK collaborations. There are external groups that produce independent catalogues where the same conclusions hold up: scatter between significance estimates.
Moreover, pipelines are not static: they are constantly developed, improved and re-configured. It is well known that the same pipeline with a different configuration can produce a different significance estimate (usually for well-understood reasons understood by the pipeline experts).
Therefore, even choosing a single pipeline can effectively represent a different pipeline per observing run (or period in which the methodology and configuration are static).
Finally, using \pastro as a threshold also utilises information from estimates of the population properties.
Since we are constantly learning new information and improving estimates, this can lead to the re-ranking of past data, resulting in the possibility of reclassifying old candidates.

One naive way of describing the situation is that significance estimates (i.e. the FAR or \pastro) do not come with an associated uncertainty (from, e.g. intrinsic configuration choices, population choices, or data choices).
The oft-used approach to resolve this is to take the scatter from multiple pipelines as a proxy indication of the uncertainty.
This has primarily been the community approach: confidence in the first detection from a new source class is validated by the involvement of multiple pipelines.
However, this is not satisfactory and discards inherent information about pipeline sensitivity.
In the remainder of this article, we will introduce a formal alternative based on \ac{CP}.
Our fundamental interest is to develop a tool that takes the \ac{FAR} or \pastro as a heuristic and calibrates it, enabling standardisation between pipelines and proper uncertainty estimates for whether a candidate is of astrophysical origin.

\section{Methodology: Quantifying Significance with Conformal Prediction}
\label{sec: cp}

We now introduce the \ac{CP} methodology.
We intend to give the reader a guide to the application without delving into the foundational theory, which can be found in reviews such as
\citet{angelopoulos2021gentle} and \citet{Shafer:2007}.

To begin, it should be understood that \ac{CP} was developed in the Machine Learning classification algorithm context.
Specifically, it can be applied to any classification algorithm, i.e. given some observed data $x$, an algorithm that produces a single predicted \emph{label} $y^{(\ell)}$ drawn from a set of $N$ possible labels $\{y^{(0)}, y^{(1)}, \ldots, y^{(N-1)}\}$.
CP calibrates the classification algorithm by producing a \emph{prediction set} $\Gamma^{\alpha}$ where $\alpha \in [0, 1]$ is the allowed \emph{error rate} also known as the \emph{significance level}.
The steps to generate the prediction set are as follows:
\begin{enumerate}
    \item  \textbf{Definitions:}
    Define a non-conformity measure $A(x, y^{(\ell)})$, which returns a non-conformity score $s$ for each label in the complete set.
    The requirements for the non-conformity score are loose; it must simply be a real-valued number.
    However, for the algorithm to be useful, the score should be large when $y^{(\ell)}$ is not the correct label (i.e. it measures how unusual the labelling would be).
    
    \item \textbf{Calibration:} Now define the calibration data: $n$ pairs of ($x$, $y^{(\hat{\ell})}$) where $x$ is the observed data and $y^{(\hat{\ell})}$ is the true label (indicated by the hat on the index).
    In our context, calibration data will always be drawn from simulations.
    Now, for each element of the calibration data, calculate the equivalent score for the true label and store this in a set of calibration scores $s_i = A(x_i, y^{(\hat{\ell})}_i)$ where the lower subscript $i$ is added to indicate the $i^{\rm th}$ element of the calibration data.
    
    \item \textbf{Quantile:} The final step before generating the prediction set is to define the allowed error rate $\alpha\in [0, 1]$, then given a set of calibration scores, we calculate
    \begin{equation}
        \hat{q} = s_{(\lceil (n + 1)(1-\alpha) \rceil)}\,,
        \label{eqn:qhat}
    \end{equation}
    where $\lceil \cdot \rceil$ is the ceiling function, and we indicate by the use of $s_{(j)}$ the $j^{\rm th}$ value of the ordered set of $s_i$.
    As described in \citet{angelopoulos2021gentle}, $\hat{q}$ is essentially the $1-\alpha$ quantile of the calibration scores with a small correction.

    \item \textbf{Prediction:} Finally, given a new observed data point $x'$, we generate the prediction set:
    \begin{equation}
        \Gamma^{\alpha} = \left\{ y^{(\ell)}: A(x', y^{(\ell)}) < \hat{q} \right\}\,,
        \label{eqn: validity}
    \end{equation}
    that is, for each label $y^{(\ell)}$, we first calculate the corresponding score $A(x', y^{(\ell)})$, then if the score is less than $\hat{q} $ we include the label in $\Gamma^\alpha$, the set of predicted labels. 
\end{enumerate}

CP guarantees that the probability that the true label is contained in $\Gamma^{\alpha}$ is approximately $1-\alpha$, this is known as \emph{marginal coverage}; more concretely, it can be shown \citep{angelopoulos2021gentle} that
\begin{align}
1-\alpha \leq \Pr (y^{(\hat{\ell)}} \in \Gamma^{\alpha}) 
\leq 1-\alpha +\frac{1}{N+1}\,,
\end{align}
such that if $N$, the number of calibration data points, is sufficiently large, we recover the standard approximate result of $1-\alpha$.

Is this useful? Practitioners in the field will no doubt know that there is a well-built-up statistical literature on decision theory behind the \ac{FAR} and \pastro introduced in \cref{sec: quant} (and we will explore this in detail in our toy model (cf. \cref{sec: cpapp}).
However, as discussed, pipelines can be miscalibrated and disagree with one another. The core motivation behind studying \ac{CP} is that we can treat the statistical quantities arising from pipelines as heuristics and use the calibration data set to adjust it, ensuring robust performance.
As we will see later in \cref{sec: extension}: this calibration process can, in fact, be viewed as a generalisation of the empirical measurement of the \ac{FAR} itself.

It is worthwhile to consider how \ac{CP} quantifies uncertainty in the label.
As scientists, we are used to talking about uncertainty on a measurement, e.g. a real-valued number accompanied by an uncertainty interval.
CP can also tackle this problem (the realm of parameter estimation or regression), but in our current context, we don't have a real-valued number; instead, we have a label.
For example, should we classify this chunk of data as containing a ``signal'' or just ``noise''?
CP provides uncertainty on the point prediction made by an underlying classifier by introducing the prediction set $\Gamma^{\alpha}$. Inspecting \cref{eqn: validity}, one can see that for binary classification of signal or noise; the four possible prediction sets are
the empty set, $\emptyset$, one of two singleton sets \{noise\}, and \{signal\}, or the double-label \{noise, signal\}.
As an anthropomorphic explanation, when asked ``does this data contain a signal or noise?'' the \ac{CP} algorithm can respond ``Neither'', ``Noise'', ``Signal'', or ``Either noise or signal''.

Varying the error rate for a fixed test data point will vary the size of the prediction set.
In the extremes: $\alpha$ close to zero or one, the \ac{CP} algorithm will be forced to respond with the double label or empty set (in the case of binary classification).
Between the extremes, the performance will depend on the problem setup and choice of non-conformity score (we will demonstrate this later).
This observation leads to the identification of what is known as the \ac{CP} \emph{confidence} \citep{Shafer:2007}, which we discuss later in \cref{sec: confidence}.

\section{Conformal Prediction for a toy Cosmic Ray detector}
\label{sec: cpapp}

We now provide a guide to \ac{CP} in the context of classification and a simple astrophysics problem: a cosmic-ray detector.
We will describe the problem and implementation qualitatively here, but the reader may wish to refer to the data release associated with this article, which contains program code to reproduce all parts of this section \citep{Ashton:2023}.

\subsection{Problem setup}
\begin{figure}
    \centering
    \includegraphics[width=8.6cm]{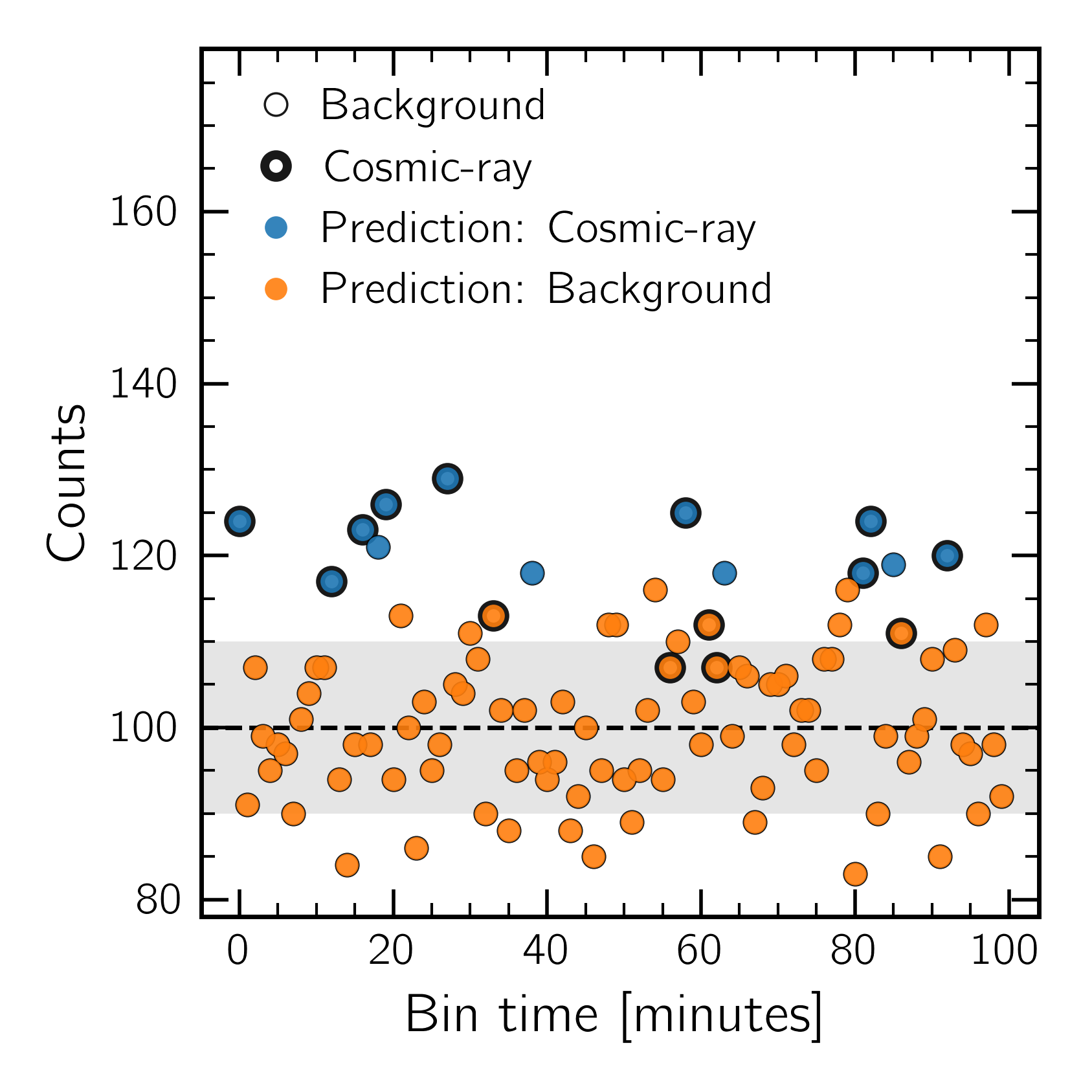}
    \caption{
    An illustrative example of data from our toy cosmic ray detector. Each data point records the number of counts within a minute-long interval or bin.
    Thick circles mark bins containing a cosmic ray.
    Data points are filled according to the prediction of the FAR detection approach: blue circles correspond to data points which surpass the threshold and, hence where we reject the null hypothesis. In contrast, orange circles indicate those that are consistent with background noise.
    }
    \label{fig: cosmic_ray_data}
\end{figure}

Consider a toy cosmic-ray detector consisting of a Geiger counter, which records the number of incidents of ionising radiation it receives per minute while pointing to the sky (this example is not intended to be realistic but indicative of typical astronomy problems).
Absent a cosmic ray, the detector will be subject to background radiation from terrestrial sources, which we model as Poisson distributed with a mean of $\lambda_{\rm b}$ counts per minute.
The detector will observe a cosmic ray as a transient burst of $N_{\rm c}$ ionising particles in some time $\delta t$, which, for the sake of this discussion, we take to be $\delta t \ll 1$~minute.
As such, we can identify and localise a cosmic ray in the data by searching for minute-long bins where the count rate exceeds the background.
The excess amount will depend on $N_{\rm c}$, which we will model again as Poisson distributed with mean $\lambda_{\rm c}$. Finally, we will also model the number of cosmic rays as Poisson distributed with some rate $\lambda_{\rm r}$ per minute.
In \cref{fig: cosmic_ray_data}, we provide an illustrative example of data from our toy detector showing minute-long bins with background, clear cosmic ray events (far above the background) and marginal cases in-between.

The standard statistical search algorithm used in cases such as this to identify if a bin contains a cosmic ray event is the frequentist one-sided $p$-value or, equivalently, the FAR.
Namely, for an observed count $c'$ and given the background rate $\lambda_{\rm b}$ 
\begin{equation}
    \FAR = \frac{1}{T}\Pr(c \ge c' | \lambda_{\rm b})=\sum_{c=c'}^{\infty} \frac{\lambda_{\rm b}^{c}e^{-\lambda_{\rm b}}}{c!}\,,
\end{equation}
where $T$ is the bin duration of 1 minute.
Note: for this toy model, we know the \ac{FAR} in closed form; this differs from the empirical FAR, \cref{eqn: FAR}, we use in gravitational-wave astronomy.

Finally, our search algorithm proceeds by applying a threshold to the $p$-value or FAR: bins above the threshold likely contain a cosmic ray, while those below do not.
In \cref{fig: cosmic_ray_data}, we apply a $p$-value threshold of 1/20 or, equivalently, a \ac{FAR} of 1 per 20 minutes.
At this threshold, we can identify four categories: several actual signals are identified (true positives: TP), but four background events above the threshold are identified as cosmic rays (false positives: FP).
Meanwhile, several cosmic rays are missed and classified as background (false negative: FN), but most background events are correctly classified as background (true negative: TN).
The non-zero counts of FP and FN are not a deficiency of the algorithm but rather inherent: with the true labels coloured in \cref{fig: cosmic_ray_data}, it is obvious which contains a cosmic ray and which does not, but our search algorithm has only the count rate leading, inevitably to errors in classification.

\begin{figure}
    \centering
    \includegraphics[width=8.6cm]{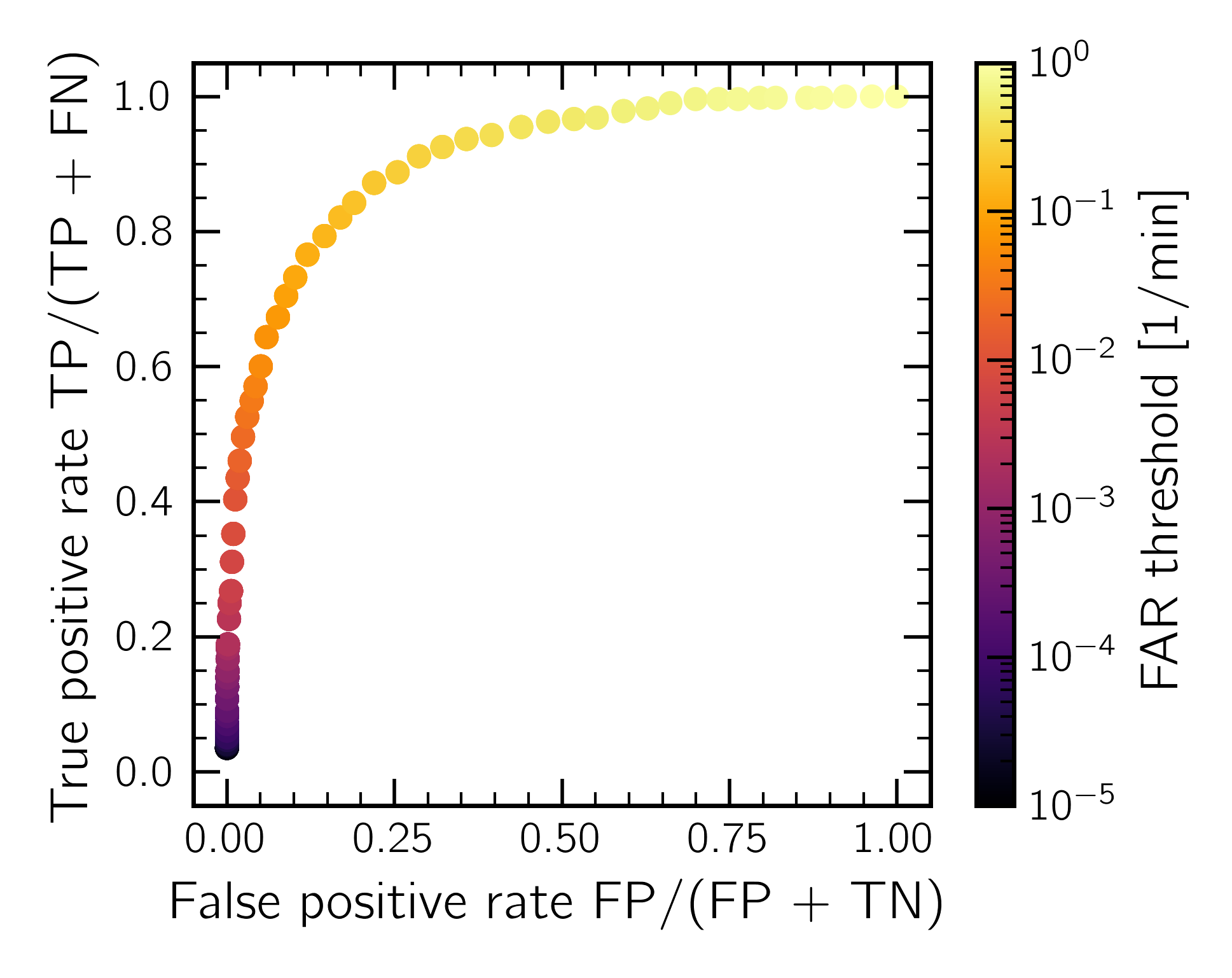}
    \caption{Measured \ac{ROC} curve for the simple cosmic-ray detector search algorithm. We measure false positive and true positive rates while varying the FAR (or equivalently, the $p$-value) threshold.}
    \label{fig: cosmic_ray_ROC}
\end{figure}

\begin{figure}
    \centering
    \includegraphics[width=8.6cm]{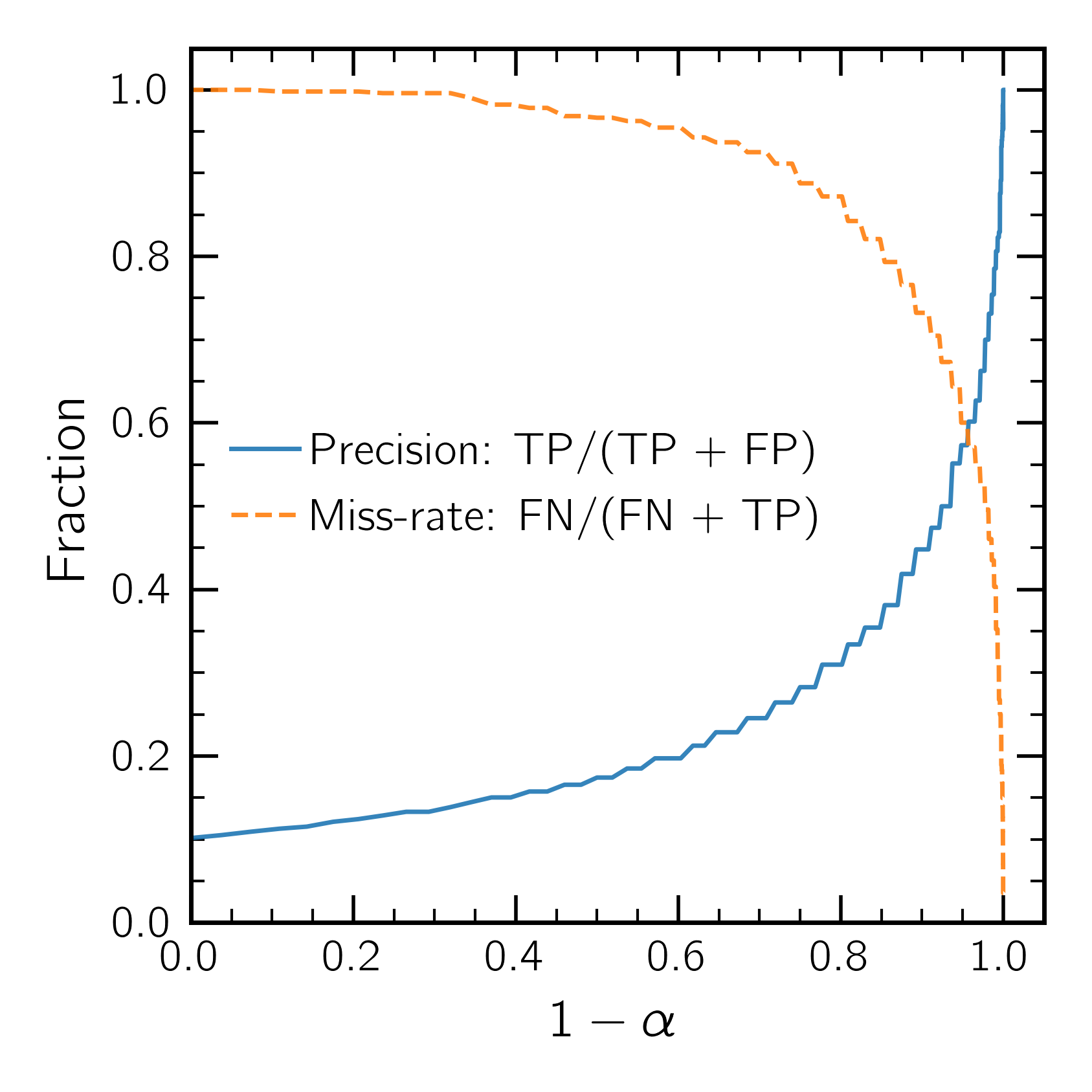}
    \caption{The precision and miss-rate for the simple cosmic-ray detector search algorithm as a function of $1-\alpha$ where $\alpha$ is the $p$-value (or equivalently the FAR).}
    \label{fig: cosmic_ray_precision}
\end{figure}

Of course, this is a well-studied problem of statistical decision theory (see, e.g., \citet{cowan1998statistical}).
In \cref{fig: cosmic_ray_ROC} and \cref{fig: cosmic_ray_precision}, we reproduce two standard figures of merit which demonstrate this behaviour.
First, the \ac{ROC} curve shows the true positive rate against the false positive rate.
The \ac{ROC} curve is generated by varying the \ac{FAR} threshold, repeatedly simulating our cosmic-ray detector, and empirically measuring the two rates.
The curve demonstrates the trade-off between true positives and false positives possible with our given search algorithm: points closer to the ideal case (top-left corner) are better in maximising the true positive rate while minimising the false positive rate.
Second, in \cref{fig: cosmic_ray_precision}, we show an alternative visualisation of the same data: the precision and miss-rate.
Considering the case of a catalogue of gravitational-wave signals, these are of more direct relevance.
The precision tells of the purity of the catalogue. If the precision is sufficiently close to 1, one can be reasonably assured the catalogue is pure and does not contain any potentially biasing terrestrial artefacts.
However, such a guarantee comes at a cost: the miss-rate tends to 0 in the same limit, indicating the catalogue size will shrink.

\subsection{Conformal Prediction}

At this point, we now step beyond the confines of classical statistical decision theory and introduce the application of conformal prediction (CP).
In this context, the cosmic-ray detector search algorithm described above can be considered a classification algorithm that produces a label $y \in \{\textrm{background}, \textrm{cosmic-ray}\}$ (whereby ``cosmic-ray'' we implicitly mean there is both a cosmic ray and background).



We apply the \ac{CP} approach defined in \cref{sec: cp} to our cosmic-ray detector problem. 
We generate a large set of calibration data points consisting of simulated data and the true classification (i.e. whether a cosmic ray was present or not).
Next, we define our non-conformity score.
We choose to use the complement of the Poisson probability mass function (noting that for the background + cosmic-ray case, the sum of two Poisson distributed variables is itself Poisson distributed with a rate equal to the sum of the rates), i.e.,
\begin{align}
    \label{eqn:cosmic_ray_non_conformity_B}
    A(x, \textrm{background}) &=  1 - \textrm{Poisson}(x, \lambda_{\rm b})\,, \\
    \label{eqn:cosmic_ray_non_conformity_C}
    A(x, \textrm{cosmic-ray}) &= 1 - \textrm{Poisson}(x, \lambda_{\rm b} + \lambda_{\rm c})\,.
\end{align}

\begin{figure}
    \centering
    \includegraphics[width=8.6cm]{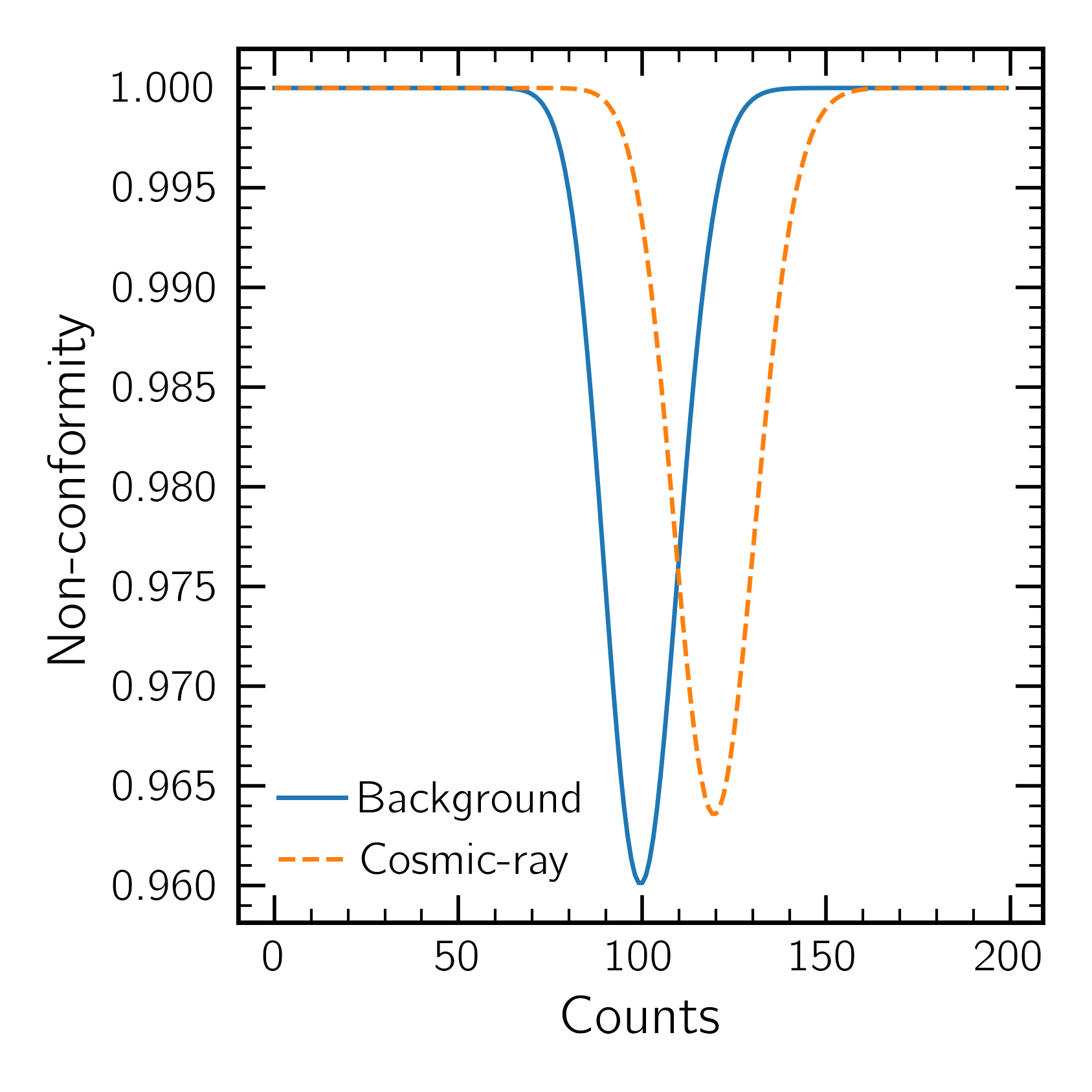}
    \caption{Visualisation of the non-conformity scores expressed in \cref{eqn:cosmic_ray_non_conformity_B} and \cref{eqn:cosmic_ray_non_conformity_C}.}
    \label{fig: cosmic_ray_nonconformity}
\end{figure}

In \cref{fig: cosmic_ray_nonconformity}, we visualise our non-conformity scores, showing that close to the mean, the non-conformity is at a minimum for each class, while away from these, they are close to unity.
We note that the absolute magnitude of the variation in non-conformity measure is not important: what matters is the relative quantile they appear when ranked by the conformal algorithm.
In this sense, the relative magnitude between classes is important (though this will not be the case later when we consider the class-conditional Mondrian conformal prediction later on).

Once our non-conformity score is defined, we can apply the conformal algorithm to new test data given some choice of $\alpha$.
For each data point, the output of the algorithm will be the prediction set $\Gamma^{\alpha}$.
In our binary case, $\Gamma^{\alpha}$ can be the empty set, $\emptyset$, one of two singleton sets \{background\}, and \{cosmic-ray\}, or the double-label \{background, cosmic-ray\}.

\begin{figure}
    \centering
    \includegraphics[width=8.6cm]{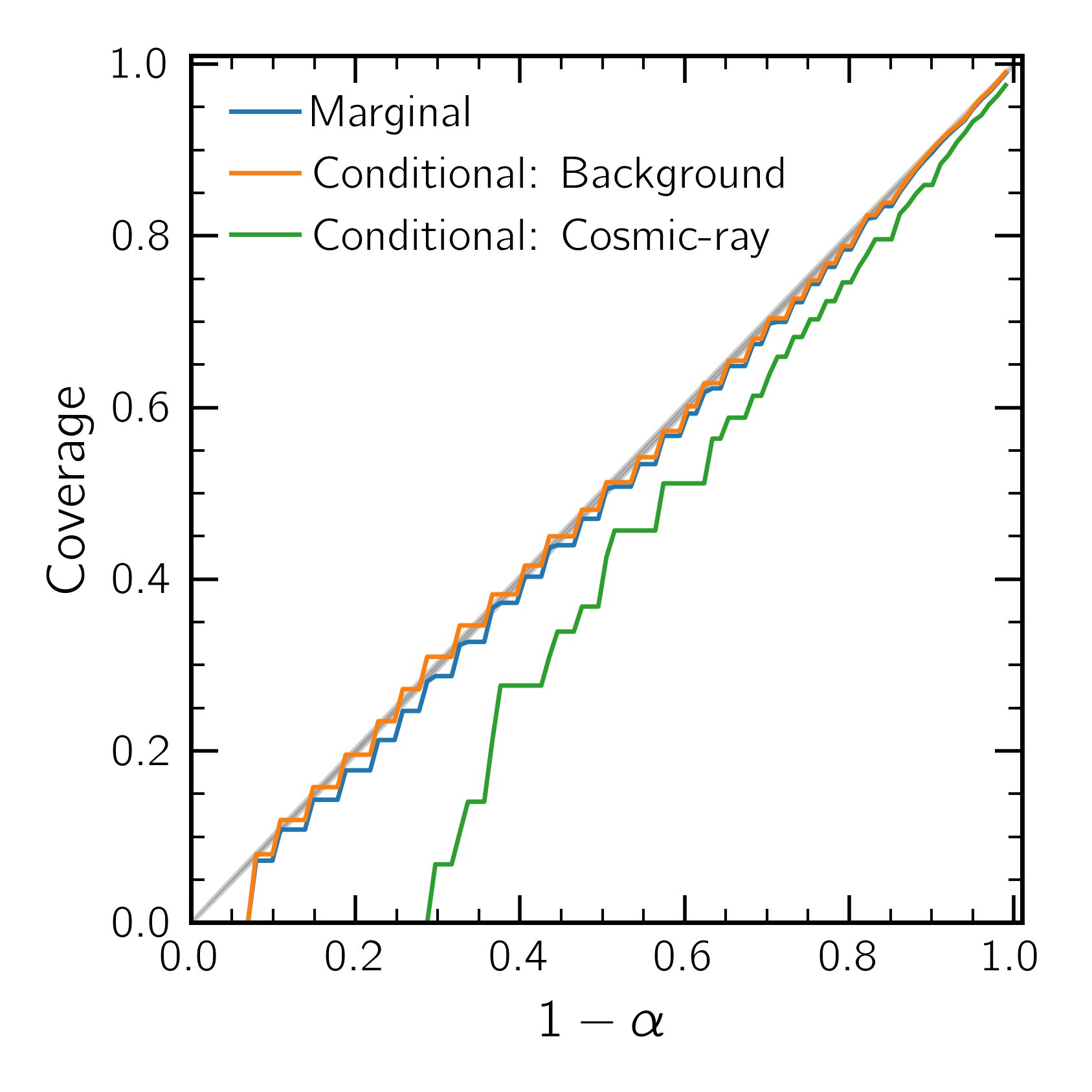}
    \caption{The empirically measured coverage (the fraction of events for which the true label is in the prediction set) for the cosmic-ray test data set after applying \ac{CP}.
    A grey band marks the 95\% binomial confidence interval expected given the size of the test data; we see variations around this due to the discrete nature of the underlying data.
    }
    \label{fig: cosmic_ray_coverage}
\end{figure}

The marginal coverage guarantee, \cref{eqn: validity}, states that, if implemented correctly, the correct label will be in $\Gamma^\alpha$ a fraction $\sim 1-\alpha$ of the time.
To check this, in \cref{fig: cosmic_ray_coverage}, we plot the empirically measured coverage after applying the conformal algorithm to a large simulated Cosmic-ray data set.
The marginal coverage (the number of times the true label appears in the prediction set) follows the one-to-one mapping guaranteed by \cref{eqn: validity}, demonstrating proper algorithm implementation.
There is some variation when $1-\alpha$ is close to zero as the set sizes become small; moreover, the step-like nature of the empirical coverage arises from the discrete nature of the Poisson data in our toy model.

\cref{fig: cosmic_ray_coverage} also provides an insight into the limitation of the simple \ac{CP} algorithm: the coverage guarantee applies only to the marginal, not the conditional labels.
As a result, the conditional labels may be over- or under-covered (i.e., exceed the allowed error rate).
We see this manifest in \cref{fig: cosmic_ray_coverage} for the cosmic-ray label, which strays away from the diagonal.
This is problematic: in gravitational-wave astronomy, we are not interested in ensuring that the label is correct as averaged over both the signal and noise labels.
We want the validity guarantee (i.e. \cref{eqn: validity}) to apply to conditional labels.
To achieve the guarantee for all labels individually, we can use \ac{MCP} \cite{vovk2013conditional}, where the data is split by class, and then the conformal prediction algorithm is applied to each group separately.
Using this technique, both the conditional labels are guaranteed to follow \cref{eqn: validity} and, by extension, the marginal labels do too.

The cost of \ac{MCP} is that the number of calibration data points entering \cref{eqn: validity} is no longer the total number but the number per label.
Therefore, the intrinsic error on rare classes consistently exceeds more common labels by design.
We apply the simple class-wise algorithm where the possible labels define the groups \citep{vovk2013conditional}.
However, more advanced approaches are possible: see \citet{ding:2023} for a formal introduction to the topic and discussion of a clustered algorithm capable of extending to many sets.

\begin{figure}
    \centering
    \includegraphics[width=8.6cm]{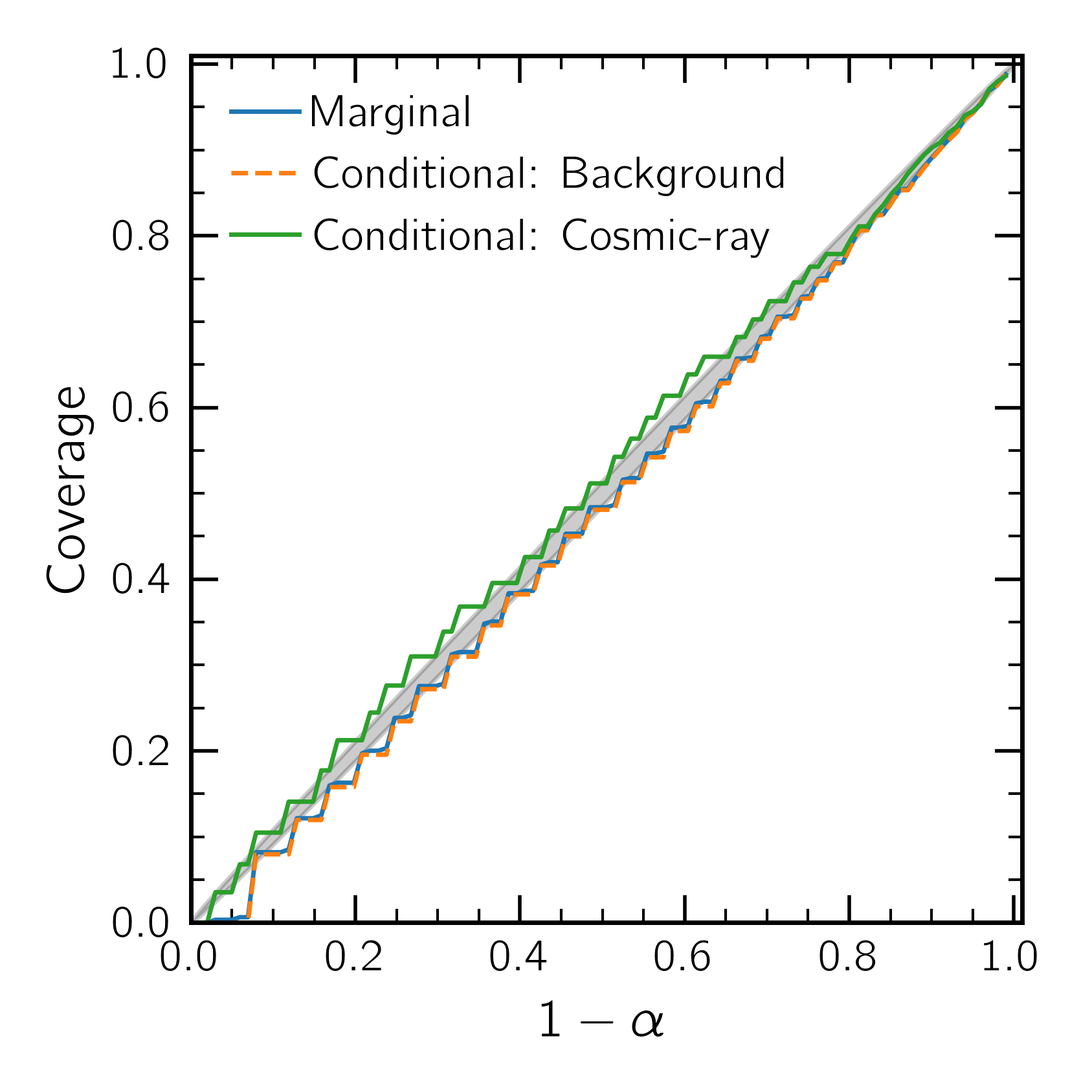}
    \caption{The empirically measured coverage (the fraction of events for which the true label is in the prediction set) for the cosmic-ray test data set after applying \ac{MCP}.
    A grey band marks the 95\% binomial confidence interval expected given the size of the cosmic-ray test data; we see variations around this due to the discrete nature of the underlying data.
    }
    \label{fig: cosmic_ray_MICP_coverage}
\end{figure}

To apply \ac{MCP}, we split our calibration data set into simulated data points containing a cosmic ray and those that do not.
Then, we apply \ac{CP} to each label and the corresponding calibration set separately for the test data.
For this reason, unlike the standard \ac{CP} algorithm, the relative values between non-conformity measures do not matter in \ac{MCP}.
In \cref{fig: cosmic_ray_MICP_coverage}, we reproduce \cref{fig: cosmic_ray_coverage} but having applied \ac{MCP}.
Now, \cref{eqn: validity} is valid for both the marginal and class-conditional labels.

\begin{figure}
    \centering
    \includegraphics[width=8.6cm]{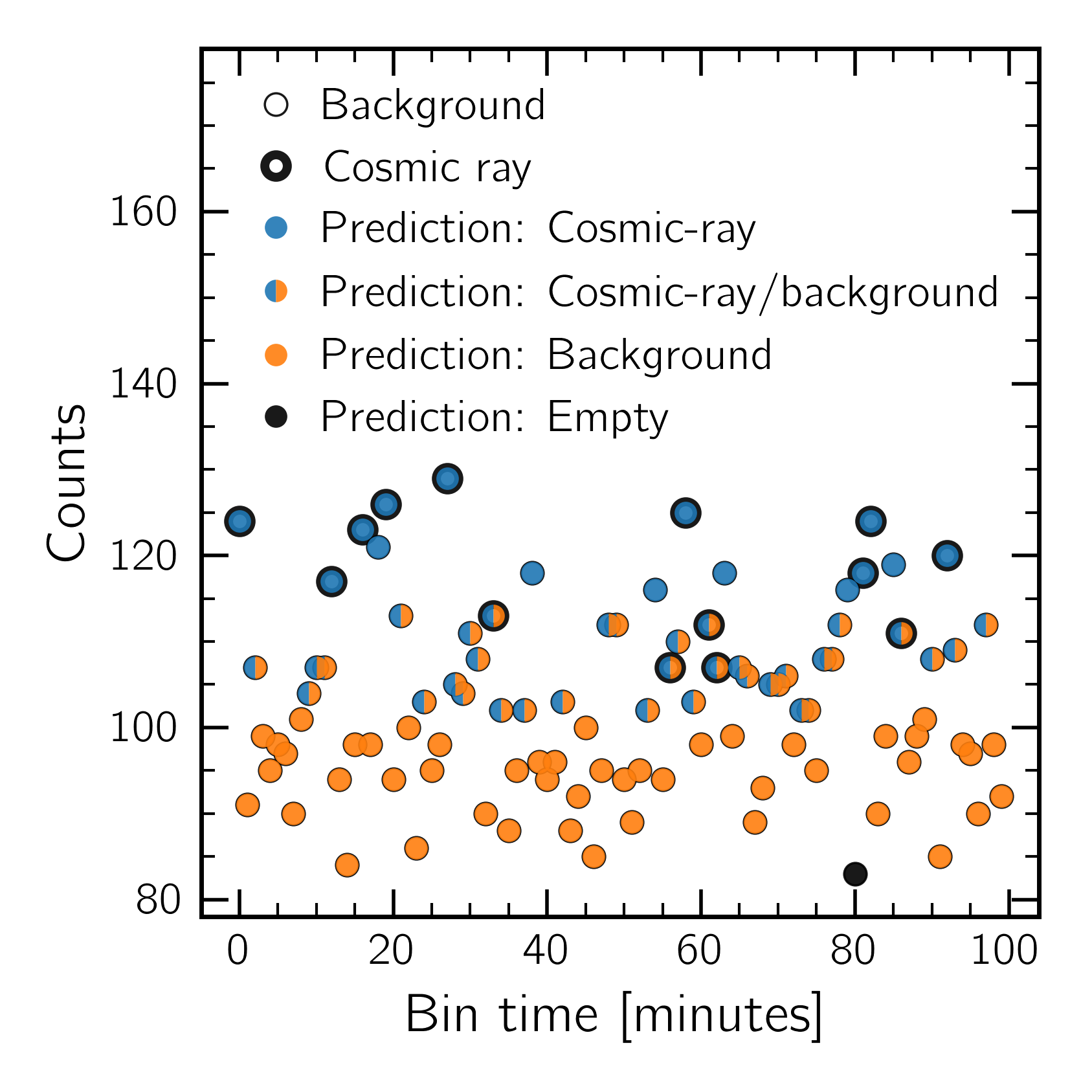}
    \caption{
    The illustrative example of data from \cref{fig: cosmic_ray_data}, but with the labels as predicted by the \ac{MCP} algorithm and using $\alpha=0.1$ (i.e. at a 90\% coverage guarantee).
    }
    \label{fig: cosmic_ray_data_MICP}
\end{figure}

\subsection{Confidence}
\label{sec: confidence}

There is a defined quantity within the \ac{CP} framework known as the \emph{confidence} \citep{Shafer:2007}.
This arises from noting that the $\Gamma^{\alpha}$ prediction sets are nested,  such that if $\alpha_1 \ge \alpha_2$, then $\Gamma^{\alpha_1} \subseteq \Gamma^{\alpha_2}$.
Since the size of $\Gamma^{\alpha}$ is a discrete quantity, it varies in steps, and these change points can be used to assign significance statements.
This observation leads us to the standard definition of confidence:
\begin{definition}
\label{def: confidence}
The confidence is the value of $\alpha$ such that the size of $\Gamma^{\alpha}$ changes from 1 to 2 (i.e. the point where we go from the single to the double label).
\end{definition}
Necessarily, each data point has a unique confidence assigned to whichever label is the \emph{single-label} given the data.

\begin{figure}
    \centering
    \includegraphics[width=8.6cm]{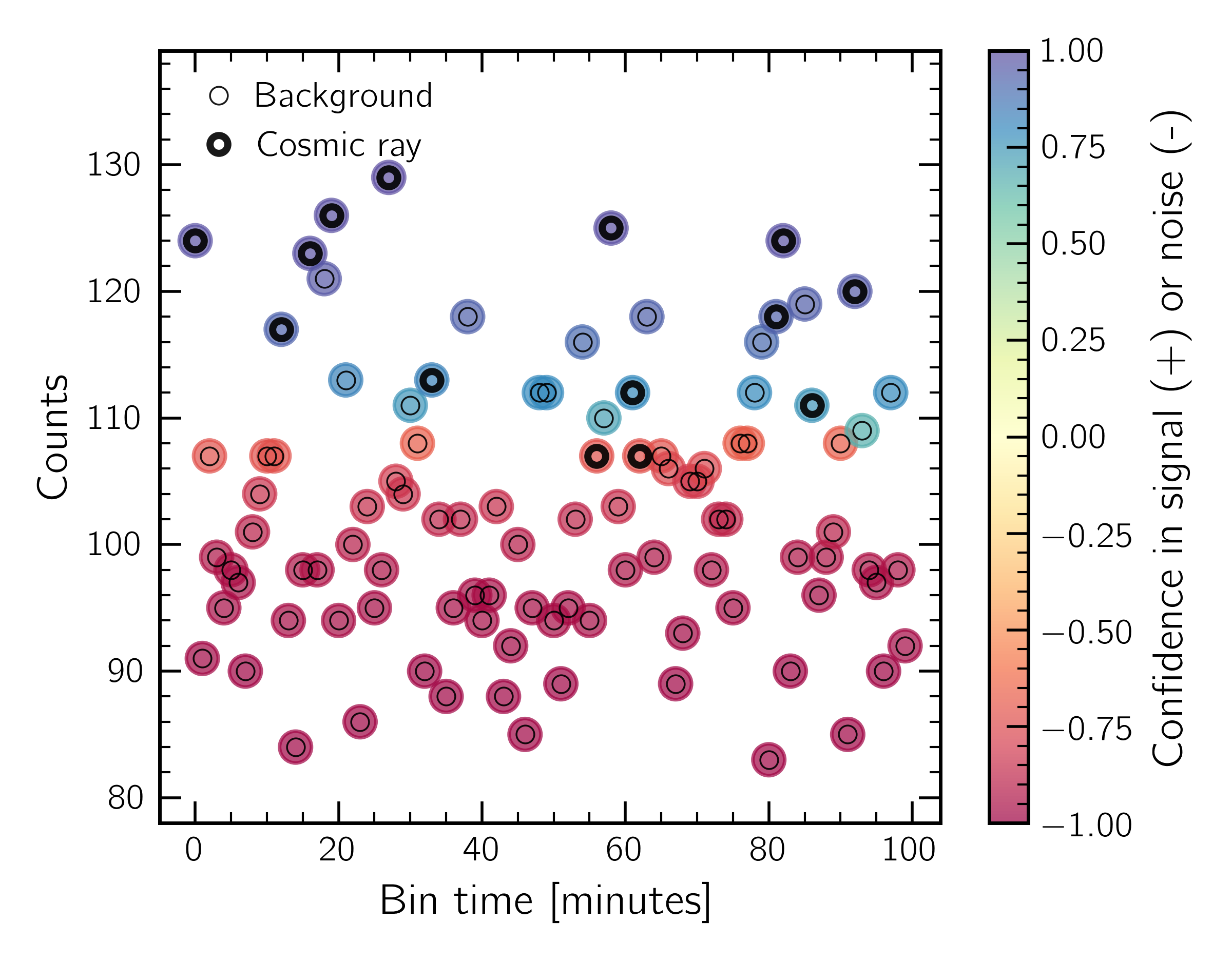}
    \caption{
    \changed{
    The illustrative example of data from \cref{fig: cosmic_ray_data} coloured by the \emph{confidence} as defined in \cref{def: confidence}. To aid visualisation, positive values are assigned to data points where the single-label prediction is ``cosmic-ray'' while we assign negative confidences to those where the single-label prediction is ``background'' (i.e. values closer to $-1$ indicate greater confidence in the noise label).
    }
    }
    \label{fig: cosmic_ray_confidence}
\end{figure}

In \cref{fig: cosmic_ray_confidence}, we take our demonstration cosmic-ray data and add the confidence, assigning $[0, 1]$ as the confidence for data points with single-label prediction ``cosmic-ray'' and flip the confidence to $[-1, 0]$ for data points with single-label prediction ``background'' (this is non-standard, but allows in the binary case to plot the confidence on a single diverging colour scale).
From this figure, we observe a sharp divide near the boundary between the non-conformity scores of the two labels (cf. \cref{fig: cosmic_ray_nonconformity}).
This notion of confidence does have uses; for example, it automatically produces a potential decision algorithm for calling something a signal: only those data points for which the single label is ``cosmic-ray''.
However, it is limited in that it does not allow one to talk about the confidence that an arbitrary data point contains a signal because, for those with a single-label ``background'', the confidence is the background confidence.

\begin{figure}
    \centering
    \includegraphics[width=8.6cm]{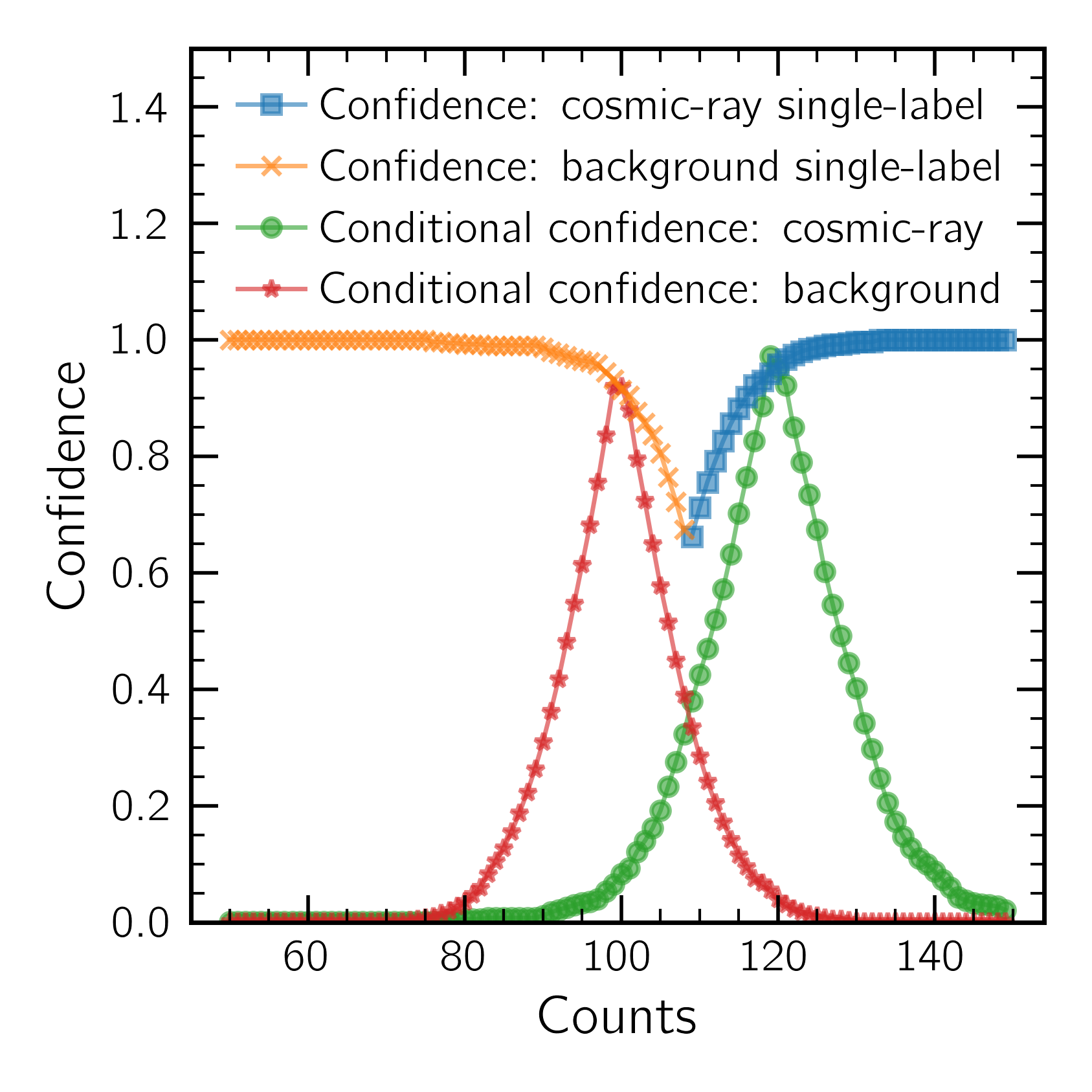}
    \caption{
    The mapping from counts to confidence (cf. \cref{def: confidence}). In blue, we show the confidence of counts where the single-label prediction is ``cosmic-ray'', in orange cases where the single-label prediction is ``background''.
    We also plot the mapping to the conditional confidence (cf. \cref{def: conditional confidence}), the cosmic-ray (green) and background (red) labels.
    }
    \label{fig: cosmic_ray_confidence_map}
\end{figure}

To further understand the confidence, we note that, in this toy example, it is a function only of the observed count rate. Therefore, as in \cref{fig: cosmic_ray_confidence_map}, we can plot the confidence as a function of the count rate to see the mapping.
In this figure, we see that at a count rate of 110 (the point where the non-conformity scores of background and cosmic-ray labels are equal, cf. \cref{fig: cosmic_ray_nonconformity}), the confidence flips between the cosmic-ray and background single label. 
There is a minimum, and on either side, the confidence monotonically increases for either label.

\begin{figure}
    \centering
    \includegraphics[width=8.6cm]{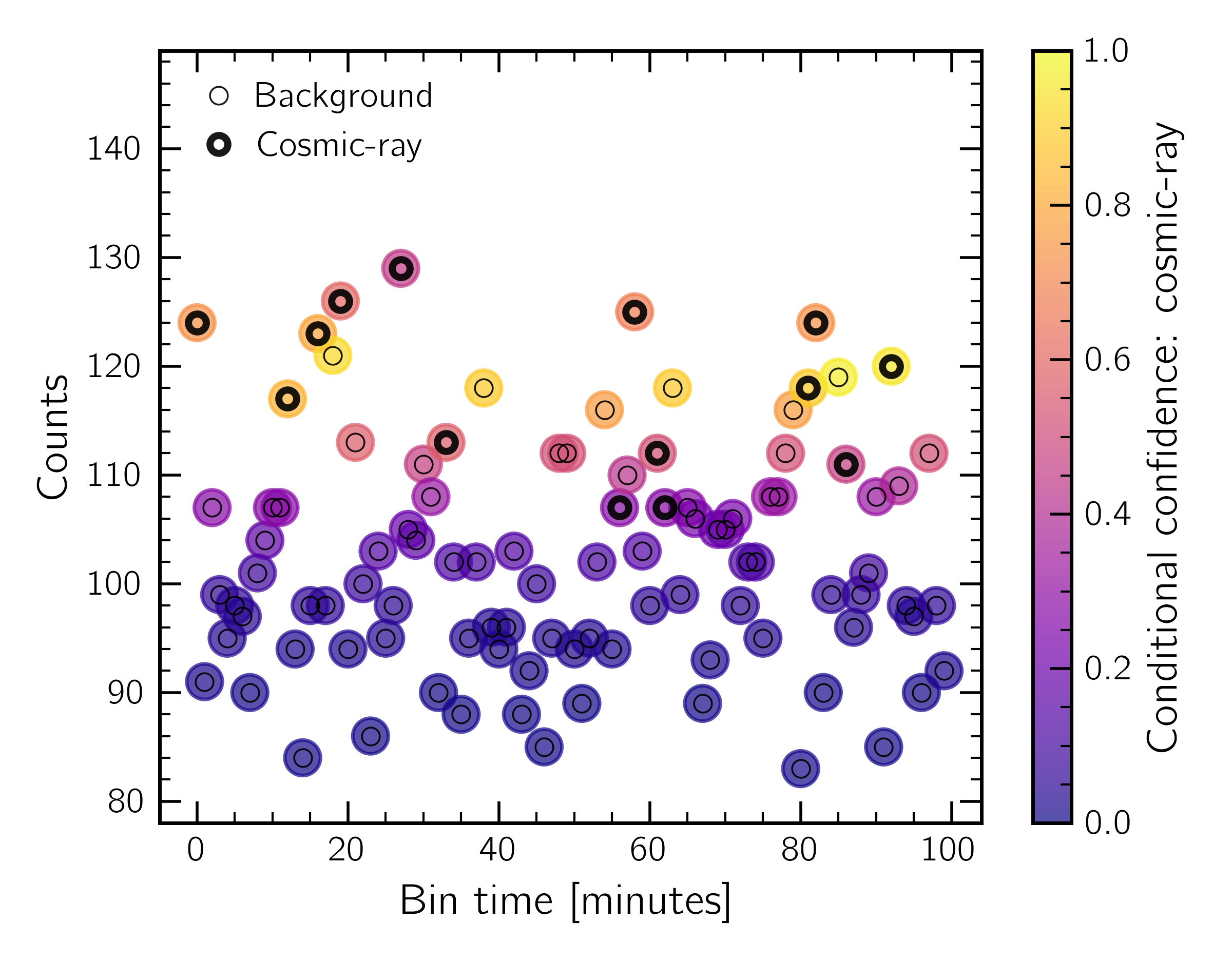}
    \caption{
    \changed{
    The illustrative example of data from \cref{fig: cosmic_ray_data} coloured by the \emph{conditional confidence} (i.e. the minimum value of $\alpha$ such that the conditional label is included in the set, cf. \cref{def: conditional confidence}) for the cosmic-ray label. 
    }
    }
    \label{fig: cosmic_ray_signal_confidence}
\end{figure}

This motivates us to consider an alternative definition, the conditional confidence:
\begin{definition}
\label{def: conditional confidence}
    The conditional confidence in label $y$ is the minimum value of $\alpha$ such that $y \in \Gamma^{\alpha}$.
\end{definition}
We add this to \cref{fig: cosmic_ray_confidence_map} for both the cosmic-ray and background labels, demonstrating that it can be calculated for any data point.
Comparing \cref{fig: cosmic_ray_confidence_map} and \cref{fig: cosmic_ray_nonconformity}, it is apparent that in this example, the conditional confidence is the scaled complement of the non-conformity score.
In a sense, this may seem circular. However, it is worth noting that the conditional confidence depends on the distribution of non-conformity scores in the calibration set and not solely on the non-conformity score itself.
Intuitively, the conditional confidence in label $y$ can be understood as the probability (interpreted as a relative frequency) that the true label is $y$ as measured from the calibration data set.
We believe conditional confidence is useful in providing an intuitive guide to understanding the significance associated with each label for a given data point.
To conclude, we finally apply the conditional confidence to our demonstration data in \cref{fig: cosmic_ray_signal_confidence} which, contrasted with \cref{fig: cosmic_ray_confidence}, demonstrates a smoother variation in assigned confidence and the ability to assign confidence in the cosmic-ray label to all data points.

\subsection{Measuring performance by set size}
\cref{fig: cosmic_ray_MICP_coverage} may give the impression that we achieved perfect performance at no cost: the calibrated \ac{CP} label sets always contain the true labels a fraction $1-\alpha$ of the time despite us never testing the performance of the conformity scores.
However, we did not consider the \emph{set size}, i.e. how many labels are given singleton labels ``cosmic-ray'' or ``background'', the double label, or no label at all?
Indeed, the set size is critical to practical utility and where we should measure the performance of our non-conformity scores.

\begin{figure}
    \centering
    \includegraphics[width=8.6cm]{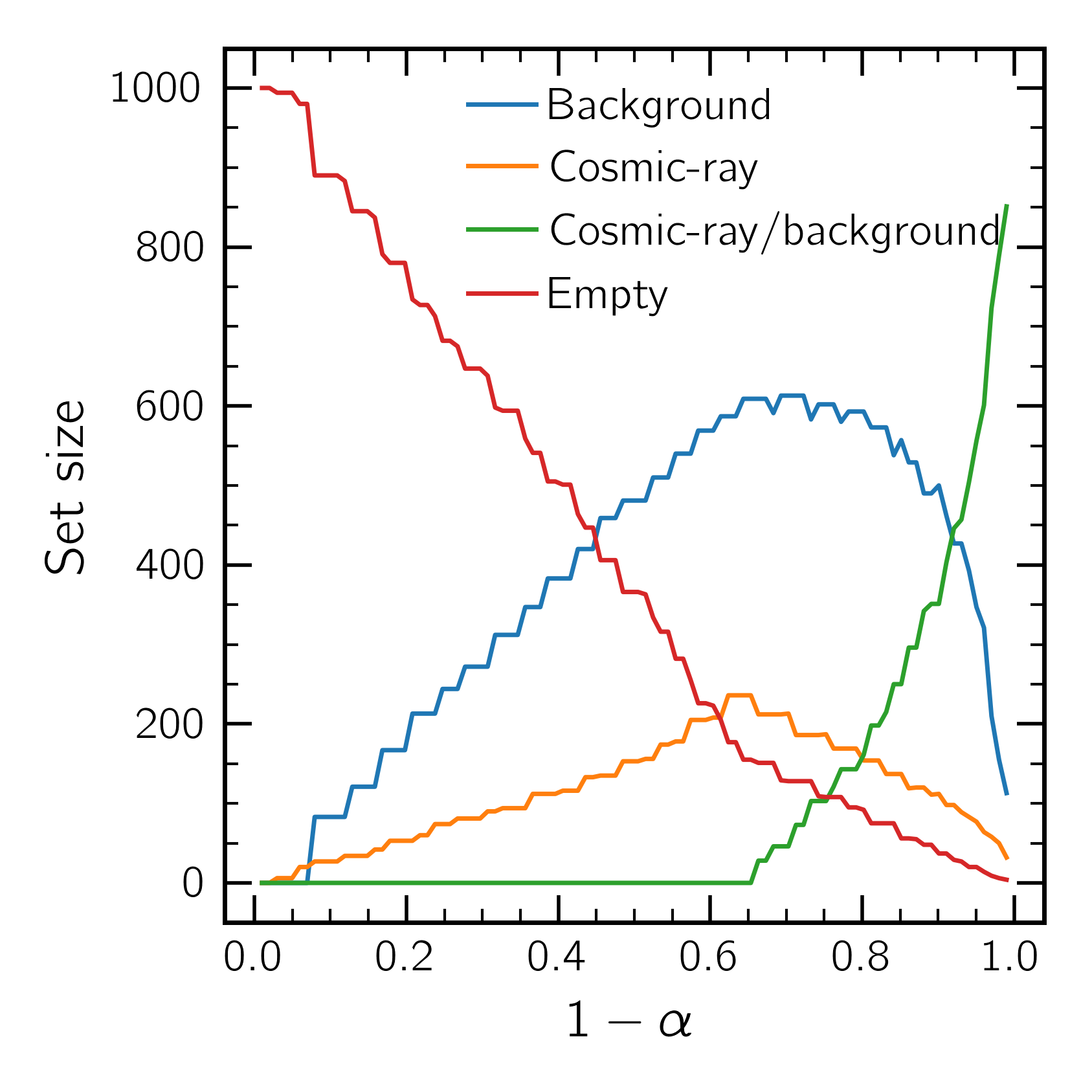}
    \caption{The set sizes for the four possible prediction sets after applying \ac{MCP} to 1000 test points for the cosmic-ray detector problem.}
    \label{fig: cosmic_ray_setsize}
\end{figure}

In \cref{fig: cosmic_ray_setsize}, we plot the set size for all four possible prediction sets as a function of $1-\alpha$.
In doing so, we show the performance: the ability to identify cosmic-ray and background events uniquely varies as a function of the allowed error rate.
At the lower extreme, we have the limiting behaviour of the algorithm. Namely, for $1-\alpha \sim 0$ (the maximum allowed error rate), all data points are in the empty set while the size of the singleton and double labels is close to zero.
For $1-\alpha \lesssim 0.6$, the set size of the singleton labels grows linearly with the size of the empty set decreasing.
Above $1-\alpha \sim 0.6$, the set size of the singletons and empty set decrease while the set size of the double label rapidly increases.

\cref{fig: cosmic_ray_setsize} explains why there is no free lunch with \ac{CP}. While we can choose $1-\alpha$ arbitrarily close to one (i.e. minimise the allowed error rate), this comes at the cost of increasing the size of the double label. I.e., the cost is a majority of triggers for which the algorithm is essentially uninformative.
Here, there is a parallel with \cref{fig: cosmic_ray_precision} in which we saw that choosing a conservative threshold increased the precision at the cost of increasing the miss-rate.
Such behaviour is unavoidable, but by measuring the set size, one can compare and optimise choices of non-conformity score.

\subsection{Performance of a poor non-conformity score}
\begin{figure}
    \centering
    \includegraphics[width=8.6cm]{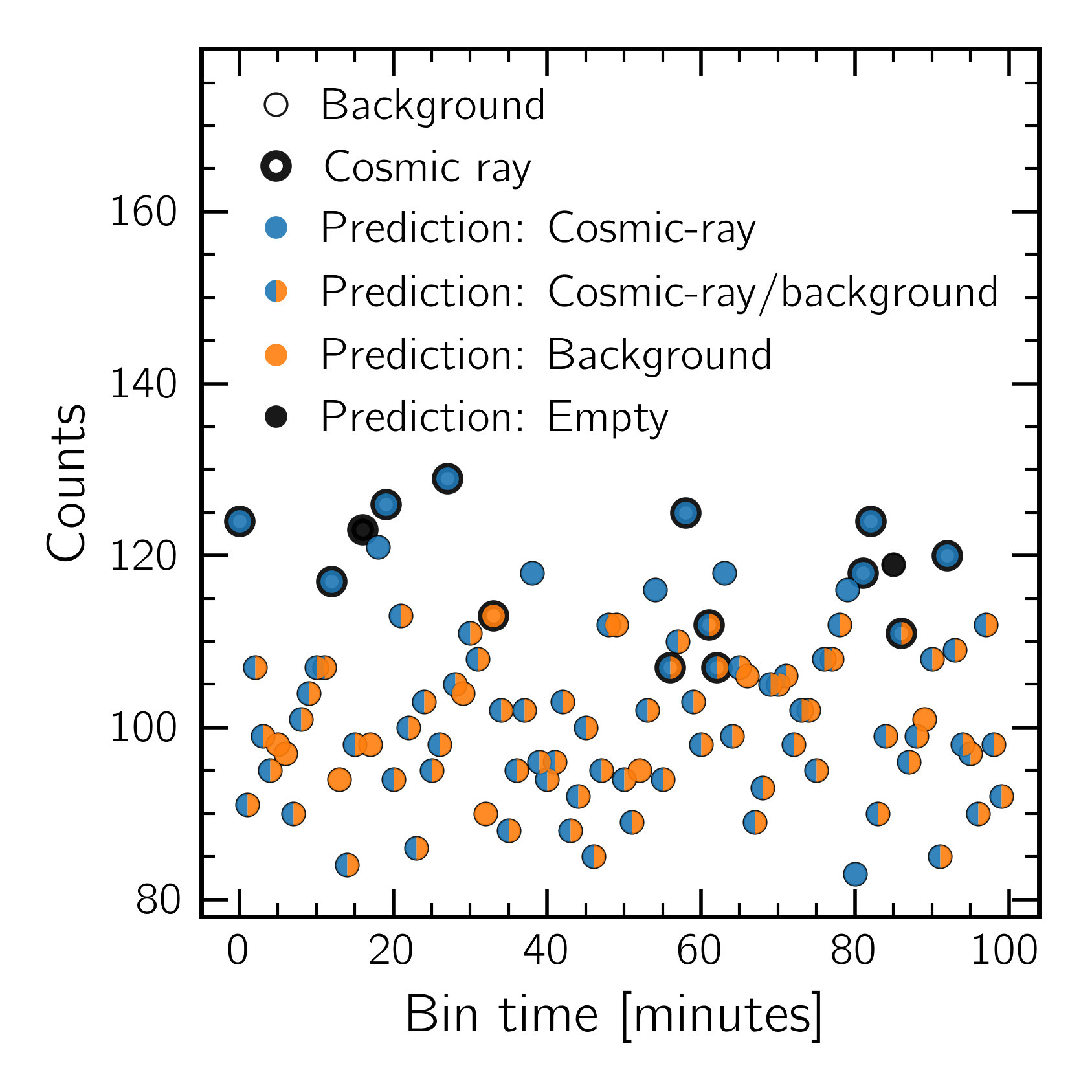}
    \caption{Reanalysis of \cref{fig: cosmic_ray_data_MICP} with $\alpha=0.1$, using \cref{eqn:cosmic_ray_non_conformity_broken_B}-\cref{eqn:cosmic_ray_non_conformity_broken_C}: a non-informative conformity measure for the cosmic-ray label. }
    \label{fig: cosmic_ray_data_broken}
\end{figure}

Finally, it is helpful to take an illustrative example of what happens when the non-conformity score performs poorly.
To demonstrate this, we take our cosmic-ray detector example and consider an alternative choice of non-conformity score:
\begin{align}
    \label{eqn:cosmic_ray_non_conformity_broken_B}
    A(x, \textrm{background}) &=  1 - \textrm{Poisson}(x, \lambda_{\rm b})\,, \\
    \label{eqn:cosmic_ray_non_conformity_broken_C}
    A(x, \textrm{cosmic-ray}) &= U(0, 1)\,,
\end{align}
i.e. while the background score stays the same, we replace the cosmic-ray score with a uniform random number generator.
We show the results by applying this to our demonstration data in \cref{fig: cosmic_ray_data_broken}.
At first, it may appear to still perform reasonably well: most of the cosmic ray events are labelled as cosmic-ray.
However, on closer inspection, we see that almost all the noise events are given the double label, multiple prominent cosmic rays have no label assigned, and background data points are labelled as cosmic rays.

This choice of the non-conformity score is extreme but yields insights into what to expect if a poor choice is made for the non-conformity score.
We can further study the behaviour by looking at the set sizes as a function of $1-\alpha$; this is done in \cref{fig: cosmic_ray_data_broken} and shows that at $1-\alpha=0.5$, labels are randomly assigned between the four choices while at either extreme either no label is assigned or the double label.

\begin{figure}
    \centering
    \includegraphics[width=8.6cm]{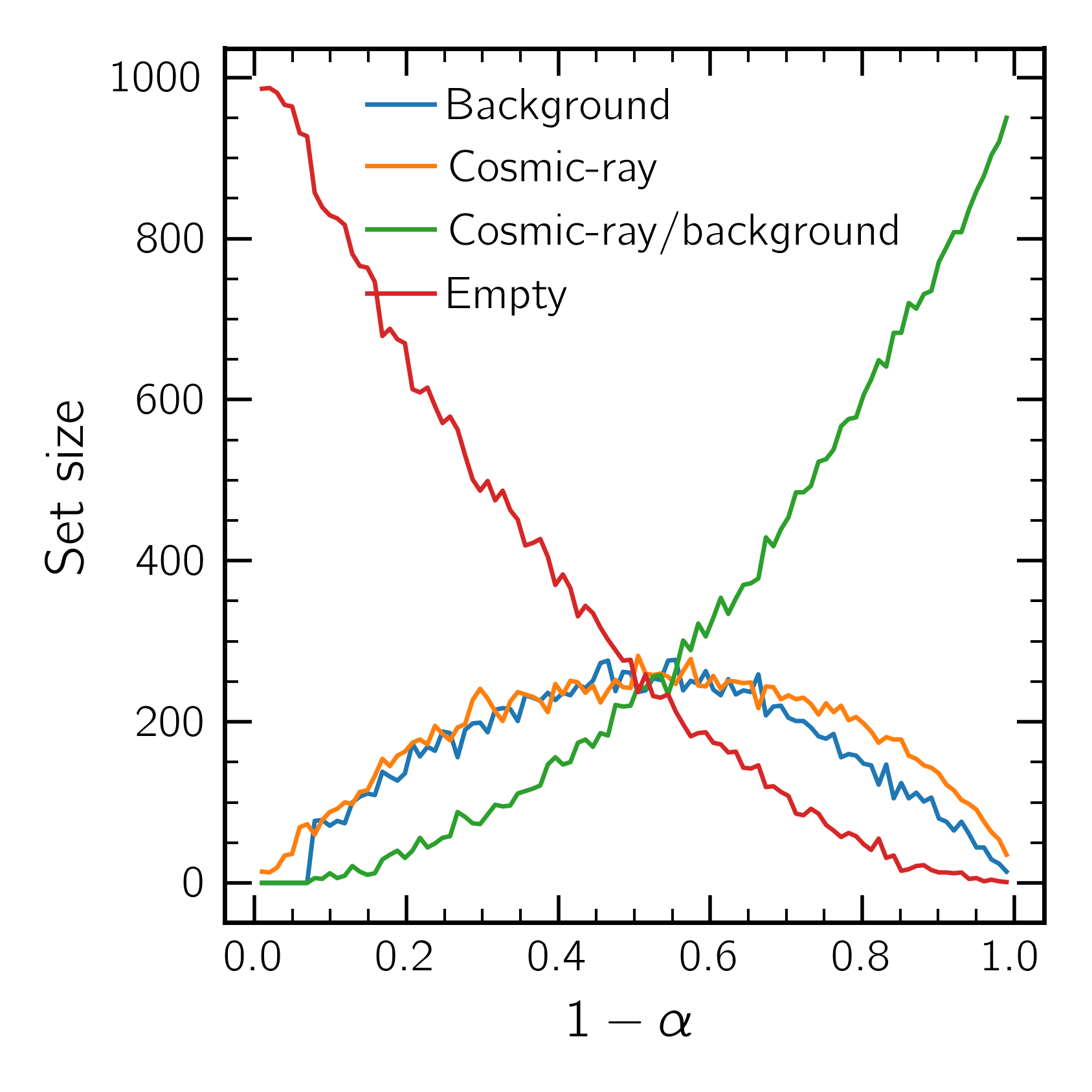}
    \caption{The set sizes of the four possible prediction sets after applying \ac{MCP} to 1000 test points with the non-conformity scores given in \cref{eqn:cosmic_ray_non_conformity_broken_B} and \cref{eqn:cosmic_ray_non_conformity_broken_C}.}
    \label{fig: cosmic_ray_set_size_broken}
\end{figure}

\changed{The set size is one way to measure the performance of a non-conformity score. For example, comparing \cref{fig: cosmic_ray_setsize} and \cref{fig: cosmic_ray_set_size_broken} we see that around $1-\alpha\sim 0.7$, the standard non-conformity score produces more single labels than either the double or background. Meanwhile, this is never true for the alternative (i.e., \cref{eqn:cosmic_ray_non_conformity_broken_B} and \cref{eqn:cosmic_ray_non_conformity_broken_C} which are intentionally broken) non-conformity scores demonstrating that the informative non-conformity measure outperformed the alternative.
The choice of non-conformity score can therefore be viewed as an optimisation problem. However, the choice of objective function is itself subjective and will depend on the use case.
For example, one option is to choose a non-conformity score that minimises the number of double labels, aiming to increase the algorithms capacity to unambiguously label the data. However, such a choice may come at the cost of increasing the empty label set. Alternatively, one may choose to maximise the TPR (or minimise the FPR) at some fixed $\alpha$.
Extending this idea, the non-conformity score itself can be parameterised, enabling direct optimisation (see, e.g. \citet{pmlr-v204-colombo23a}).
Regardless of the methodology, the choice of objective function for the optimization will always be subjective and the best choice will depend on the
overarching use case. For gravitational-wave astronomy, we anticipate some combination of maximising the number of single labels while minimising the number of false positives, but we intend to explore this in future work.
}

\section{Conclusion: toy model}
In this section, we have used a simplistic toy model to introduce \ac{CP}.
In the main, we use this as a tool to understand CP and not as a demonstration of the application of CP to realistic astrophysical problems.
We recognise that there are steps that do not transcend, e.g. here, we know the statistical properties of the signal and noise distributions perfectly and can use these to construct a non-conformity score.
Nevertheless, we hope it may prove useful as a starting point for others to apply \ac{CP} using the accompanying notebook \citep{Ashton:2023}.

\section{Conformal Prediction for gravitational-wave astronomy}
\label{sec: cbc}

Having introduced \ac{CP} for a simple toy model, we now extend the discussion to gravitational-wave astronomy. 
We will focus on the use case of modelled transient searches for \ac{CBC} signals. However, the discussion applies generally since the standard statistical framework is applied across the field.

Our primary task is to define the non-conformity measure $A(x, y)$.
Considering the binary classification problem, signal or noise, two obvious initial choices exist: using the \ac{FAR} or the Bayesian \pastro quantities.
For source classification, e.g. binary black hole, neutron star black hole, binary neutron star, or terrestrial, one could use the multi-class \ac{CP} algorithm and the Bayesian probabilities provided by the pipeline for each source class.
Therefore, these choices are readily applied to the outputs of existing pipelines, which is what we choose to do in this work.

However, \ac{CP} offers scope for further development.
For example, the \ac{FAR} used by pipelines uses a ranking statistic combining the matched-filter \ac{SNR} and $\chi^2$ statistic amongst other quantities.
Such a combined ranking statistic can itself be used as a non-conformity score: in effect, the ``calibration'' data set of \ac{CP} is then analogous to the background data used in a traditional search pipeline.
Building on this idea, if the combination is parameterised, one could optimise the ranking statistic (non-conformity score) to minimise the counts of the empty set of multi-label prediction sets on some test data.
Such an idea builds on a similar application by \citet{McIsaac:2022odb}, which seeks to maximise the separation of signals and noise.
Many more such innovations are likely possible.

\subsection{Using Conformal prediction to calibrate multiple competing pipelines}
\label{sec: applying}

To demonstrate the application of \ac{CP} to gravitational-wave astronomy, we will use the results of a recent \ac{MDC} study in advance of the LVK fourth observing run \citep{Chaudhary:2023vec}.
In this \ac{MDC}, four low-latency \ac{CBC} online search algorithms were applied to a real-time data replay from the third observing run.
Simulated signals were added to the data at a rate much greater than the anticipated astrophysical rate under current detector sensitivities.
This higher rate was used to stress-test the low-latency infrastructure: the primary goal of the \ac{MDC} was to measure expected performance in producing public alerts used to trigger event follow-up. 
Taking the MDC data, we adjust classifications for all real gravitational-wave detector events present in the \ac{MDC}, but do note there are potential sub-threshold signals that remain.
We also remove all early warning triggers from the \ac{MDC} and use the corrected \pastro values from \citet{Ray:2023nhx}.

The \ac{MDC} data products provide a perfect test bed for \ac{CP}.
The increased rate produces a sizeable set of simulated triggers, e.g. points in the data stream that the search pipelines identify as likely to contain a signal.
Most recorded triggers in the \ac{MDC} are simulated signals (this differs from the astrophysical scenario where, at a high \ac{FAR} threshold, most triggers will be non-astrophysical noise).
Moreover, the configuration of the pipelines was in development during the \ac{MDC}, leading to imperfect performance.
For these reasons, the performance of the pipelines is not representative of the tuned performance expected during the run.
This point is discussed within \citet{Chaudhary:2023vec} specifically for the case of \pycbc: ``
The FAR values for injections recovered during the MDC are subject to a substantial upward bias due to the high rate of high-\ac{SNR} injected events, which significantly influences the background estimation.''
As a result, in the context of candidate significance estimation, we can consider the \ac{MDC} data as the application of poorly calibrated pipelines to a given data set.
It, therefore, is a good test bed to show how \ac{CP} can automatically calibrate the pipelines.
However, we stress that the following discussion should not be taken as indicative of the performance of the pipelines, only as an example where they are known to be ill-tuned.

\begin{figure}[t]
    \centering
    \includegraphics[width=8.6cm]{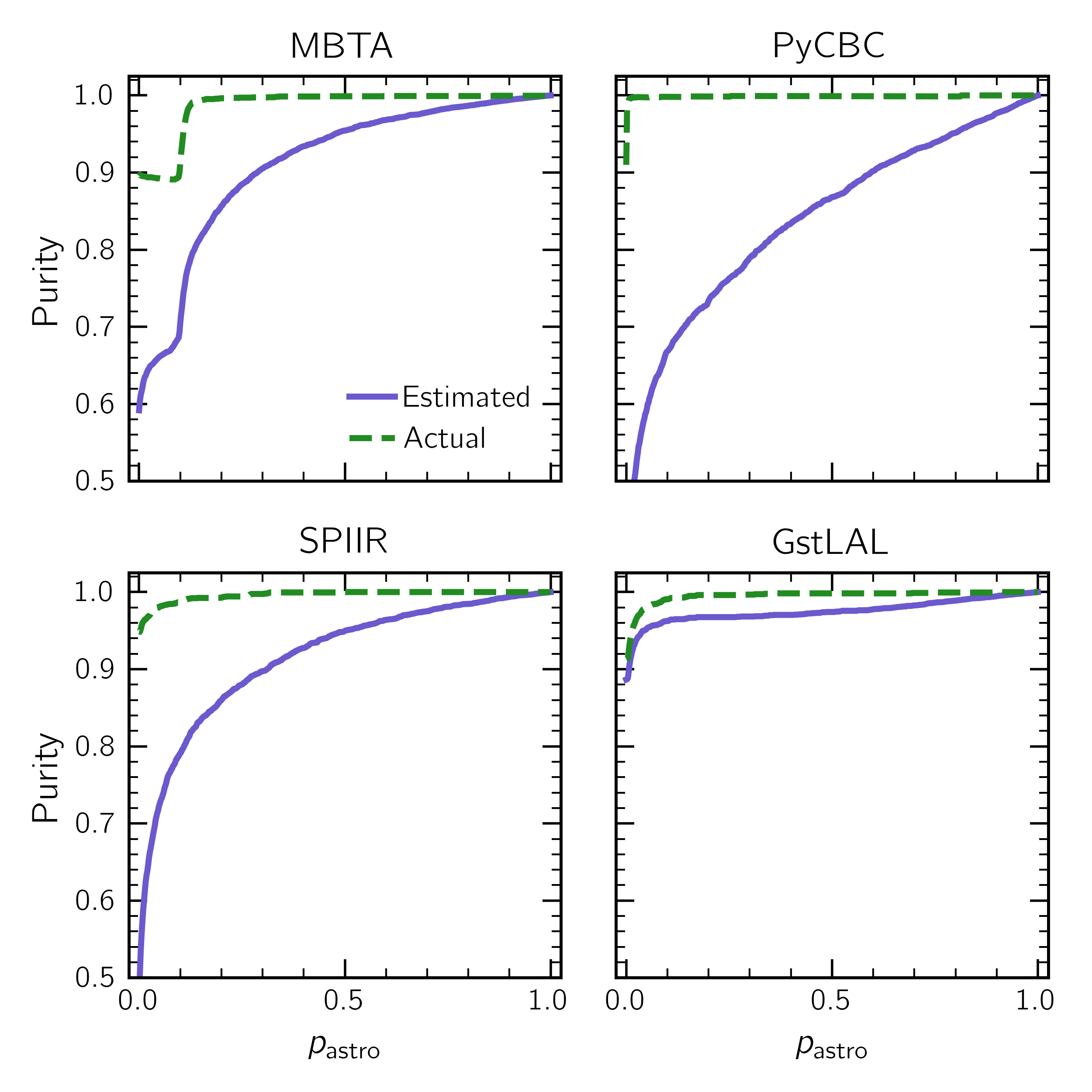}
    \caption{The estimated and actual purity for the \ac{MDC} results as a function of the \pastro threshold split by pipeline.
    Estimated purity refers to the sum of \pastro above the threshold, while actual purity refers to the count of triggers pertaining to signals above the threshold.
    We use purity here as it is the common language of the field. However, we note that it is identical to the coverage defined in the field of \ac{CP}.}
    \label{fig: MDC11_pastro_coverage_bypipeline}
\end{figure}

Let us begin by studying the performance of the pipelines using traditional significant estimation approaches.
We start by thinking about the catalogue of events that would be produced at a given threshold.
In \cref{fig: MDC11_pastro_coverage_bypipeline}, we plot the purity of the resulting catalogue as a function of \pastro; we present results separated by pipeline.
We calculate the purity as the fraction of triggers with \pastro greater than the threshold which pertains to an injected signal.
We plot the actual purity (the true number of simulated signals in the trigger set) and the estimated purity: the sum of the \pastro for all triggers above the threshold.
The sum of \pastro to estimate the number of astrophysical signals is commonly used in the context of a catalogue of triggers (see, e.g. \citet{GWTC3}). It formally amounts to the posterior-estimated number of foreground events in the \citet{Farr:2013yna} framework.
\cref{fig: MDC11_pastro_coverage_bypipeline} shows varying behaviour by pipeline, with all pipelines under-estimating the actual purity by varying amounts.
(We note that, due to the presence of potential sub-threshold real signals in the MDC data, the ``Actual'' estimate here is potentially biased; however, given the expected purity of sub-threshold candidates in GWTC-3 \citep{GWTC3}, the level of bias is at most a few percent).
By comparison, the advantage of \ac{CP} is that $\alpha$, the allowed error rate of the algorithm, maps directly onto the actual purity of the resulting catalogue.

To demonstrate \ac{CP} in practice, for the set of candidates from each pipeline, we evenly split the \ac{MDC} data results into a calibration and test set.
We then apply \ac{MCP} using the \ac{FAR} as the non-conformity score for `signal' and the \ac{iFAR} as the non-conformity score for `noise'.
This way, we use the pipeline outputs directly without adding additional information.
We then apply \ac{MCP} to each trigger in the test data set, using the calibration data for producing a prediction set.
Note that the computational effort required for this step is negligible (a few CPU seconds on any modern computer).

\begin{figure}[t]
    \centering
    \includegraphics[width=8.6cm]{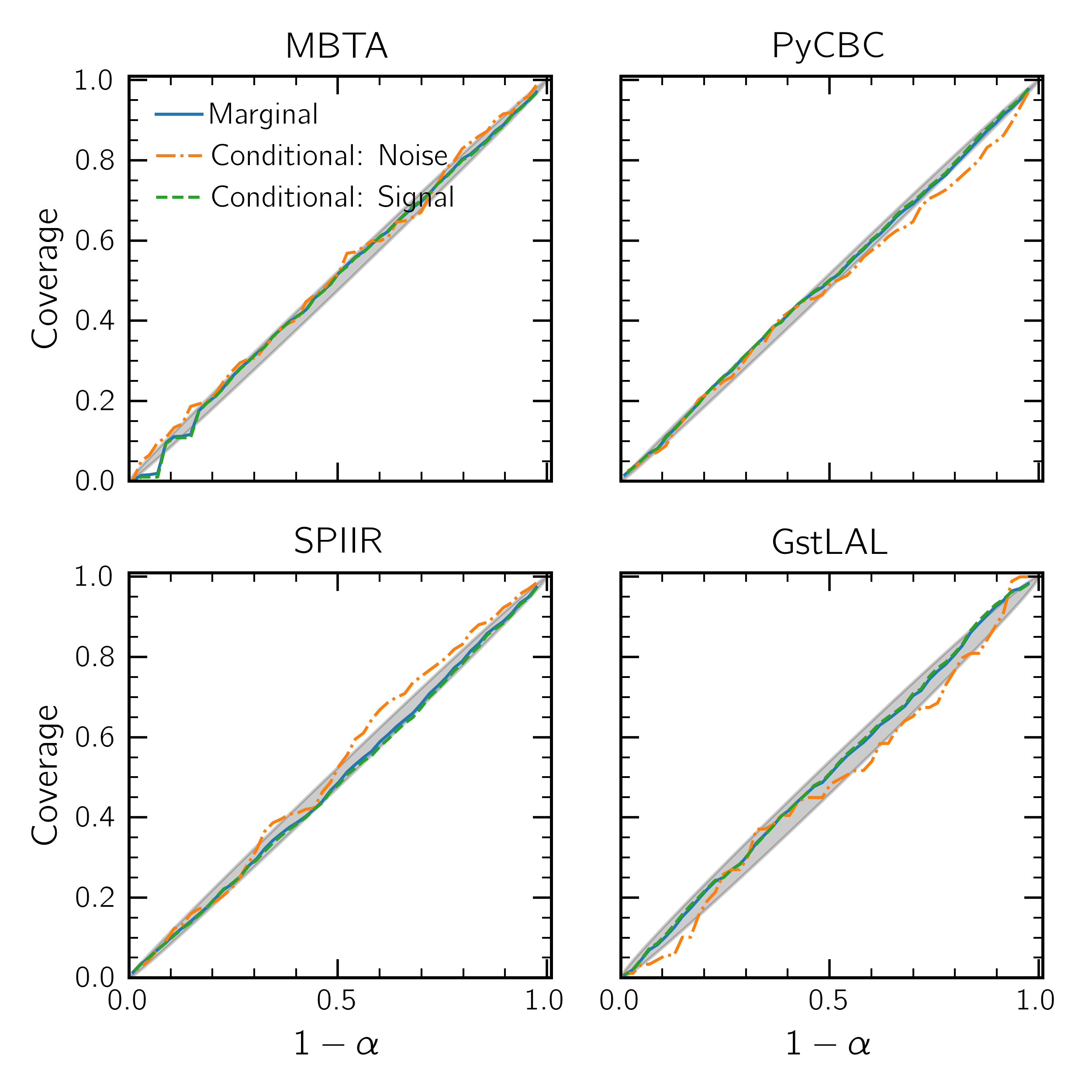}
    \caption{
    The marginal and conditional coverage for all \ac{MDC} results after applying \ac{MCP}, demonstrating they satisfy the validity guarantee.
    \changed{Recall that the marginal coverage is averaged over all labels, while conditional coverage is as applied to a single label at a time}.
    A grey band marks the 95\% binomial confidence interval expected, given the size of the entire test data for each pipeline. Note that for the conditional labels, the size of the effective test data set is smaller, and therefore, the anticipated Poisson counting error can be larger, as is the case of the \gstlal conditional noise label.
    }
    \label{fig: MDC11_error_coverage_bypipeline}
\end{figure}

In \cref{fig: MDC11_error_coverage_bypipeline}, we plot the label coverage for each pipeline, demonstrating it satisfies \cref{eqn: validity}, i.e. for all $\alpha$, the fraction of test triggers which contain a simulated signal has a one-to-one correspondence with $1-\alpha$.
\changed{Moreover, we note that all pipelines satisfy this: irrespective of their underlying performance, once calibrated by \ac{CP} the coverage guarantee is ensured.}
We now note that what is known in the field of \ac{CP} as coverage is equivalent to the catalogue purity.
As such, \cref{fig: MDC11_error_coverage_bypipeline} and \cref{fig: MDC11_pastro_coverage_bypipeline} can be contrasted to show how calibrating with \ac{CP} regularises the meaning of the threshold between pipelines.
The implication is that once calibrated by \ac{CP}, the catalogue produced at a fixed $\alpha$~threshold contains an a priori known contamination rate:~$\alpha$.
Therefore, downstream analysis can decide the contamination rate they are willing to accept and then use that to set the threshold for inclusion.

\subsection{Understanding individual events: confidence}
In the last sub-section, we saw how a catalogue could be created by applying \ac{MCP} to calibrate the significance estimates.
Such an application guarantees the purity of the resulting catalogue. It is, therefore, directly applicable to the case of population analyses, where one often needs to control the purity over a set of triggers.
However, this leaves the question of assessing individual events and deciding if they are astrophysical, which we now discuss.

In the traditional framework, candidate significance is assessed by combining the FAR, \pastro, their constituent elements (e.g. the $\chi^{2}$ statistic), and a deep knowledge of the performance of the pipeline.
For example, the first direct observation of gravitational waves from GW150914 \citep{GW150914} reported a \ac{FAR} of 1 event per 203,000 years (and gave an equivalent $>5\sigma$ estimate).
However, once a source class is established, \pastro is generally the preferred mechanism to identify new events (for example, independent reanalyses use this criterion \citet{Venumadhav:2019lyq}).
However, for newly detected source classes, because \pastro requires an astrophysical model of the rates, which is generally poorly constrained, it is common to revert to a more detailed study of the \ac{FAR} (see, e.g. the discovery of the first neutron star black hole mergers \citep{NSBH}).

\begin{figure*}
    \centering
    \includegraphics[width=5.8cm]{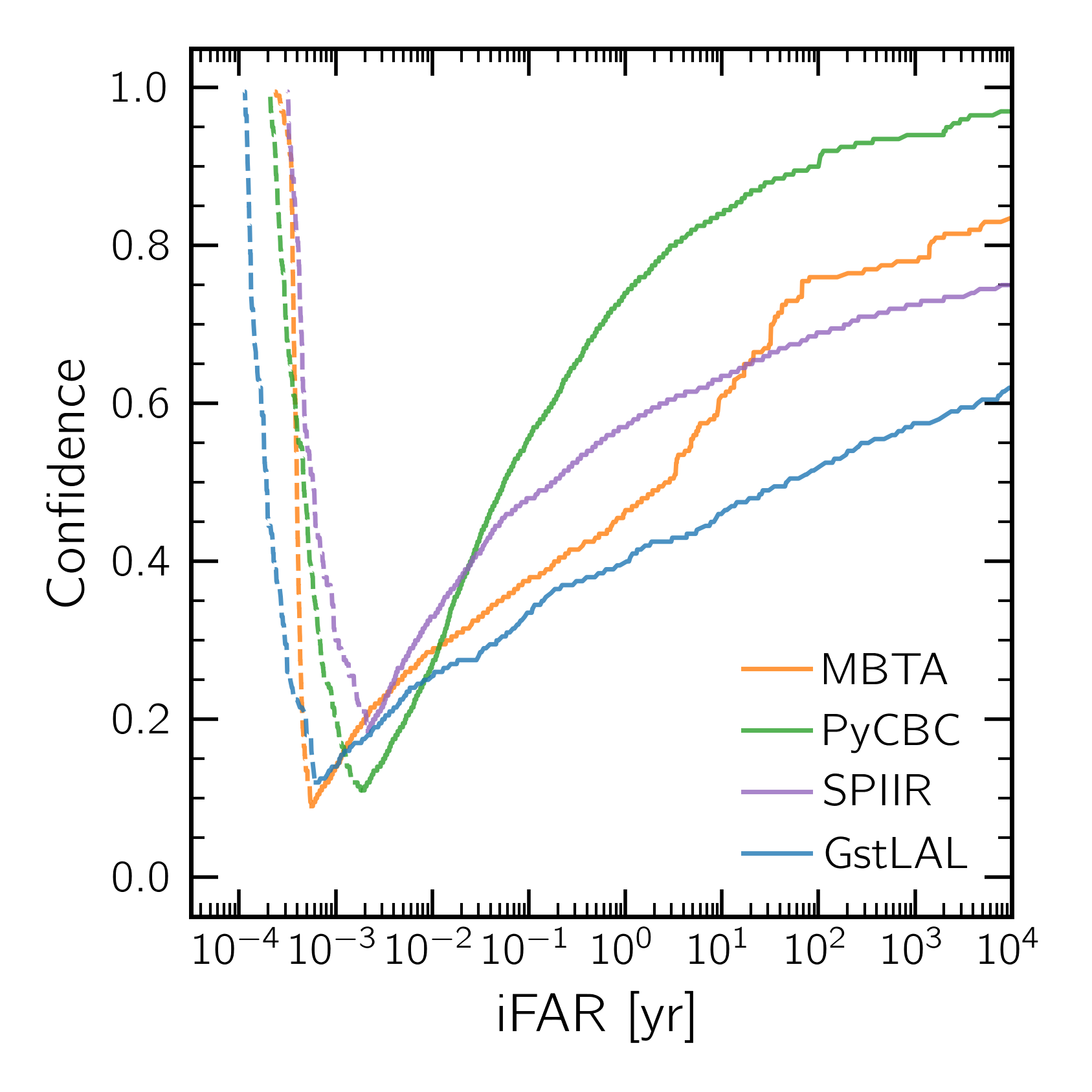} 
    \includegraphics[width=5.8cm]{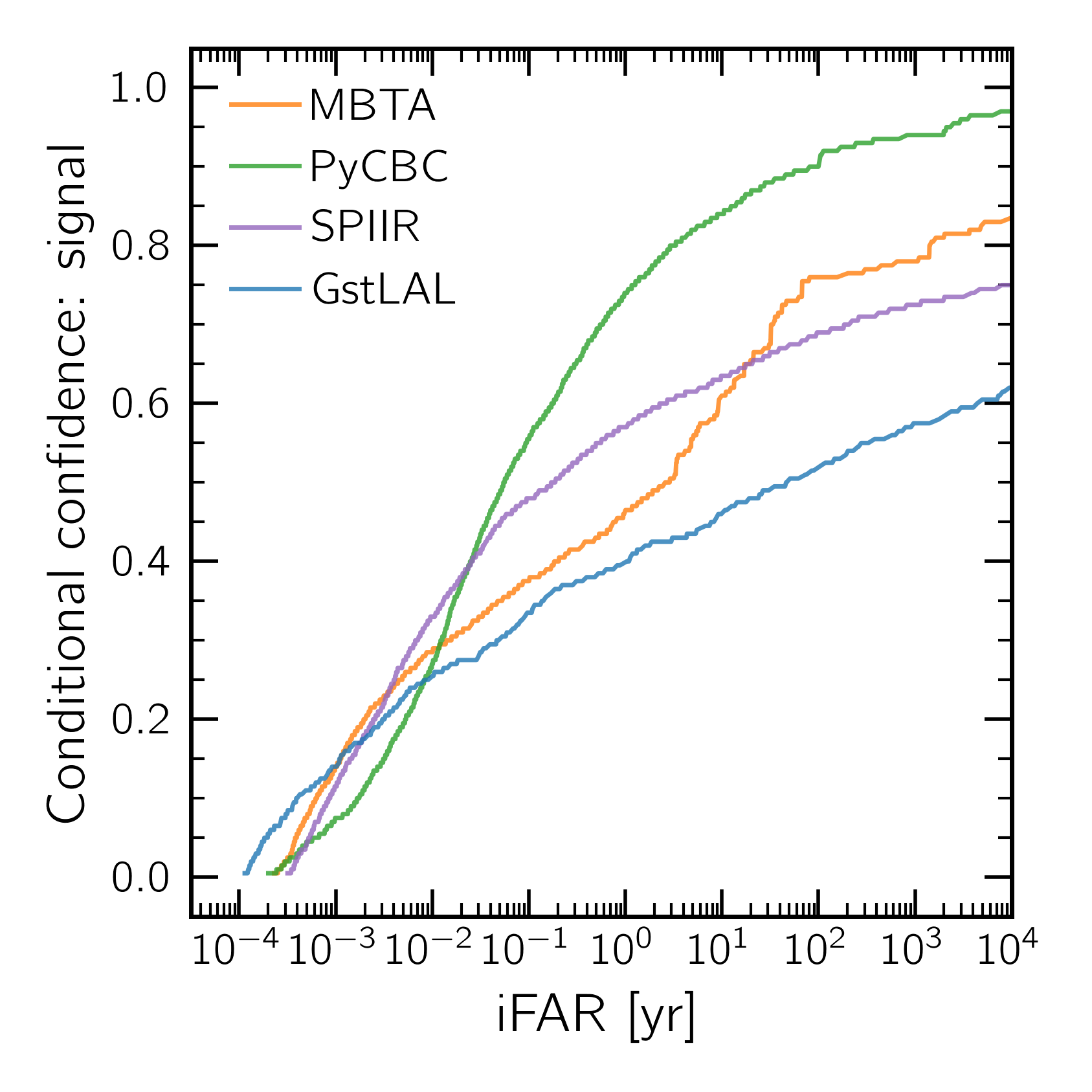}
    \includegraphics[width=5.8cm]{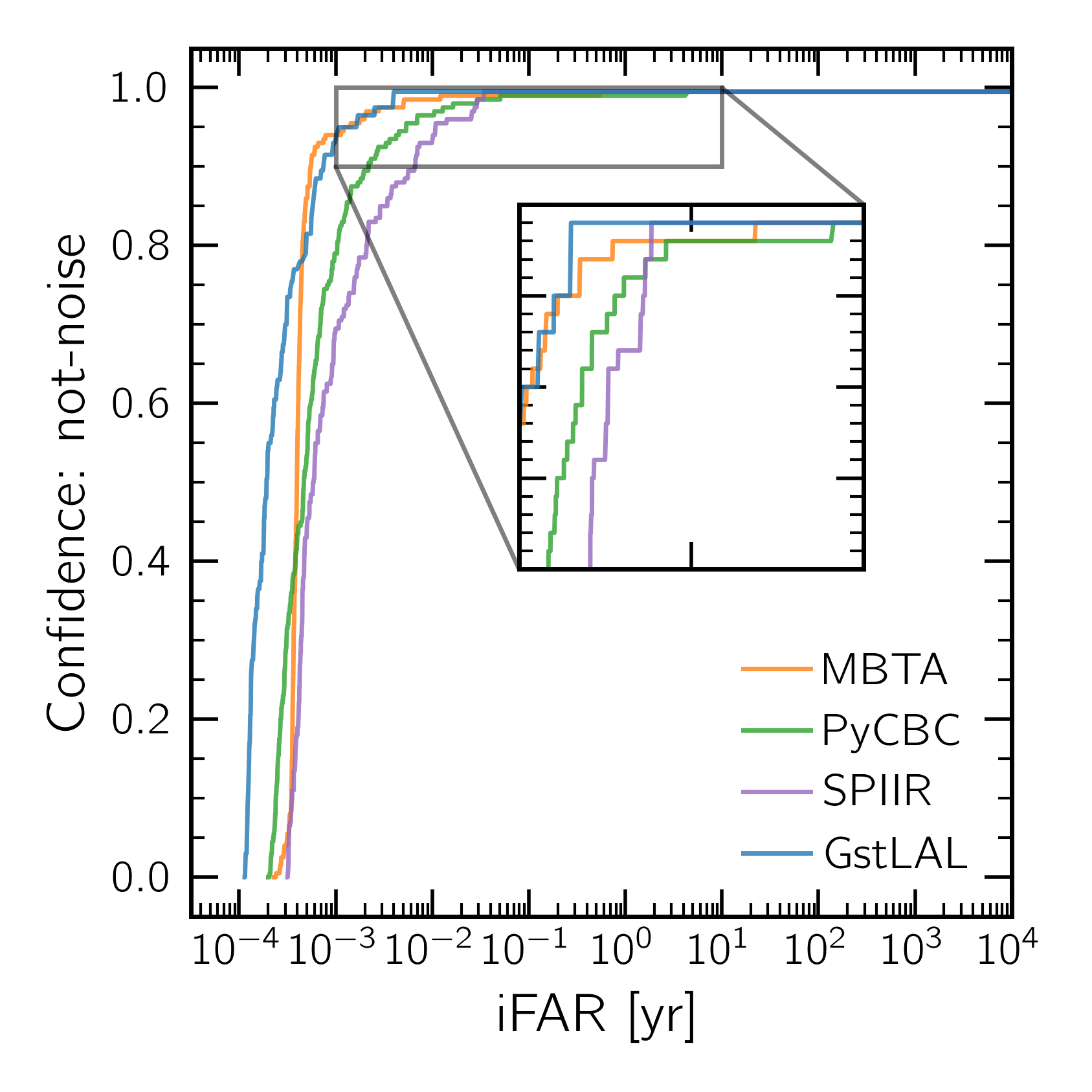}
    \caption{
    The relation between the \ac{iFAR} and three definitions of \emph{confidence} within the \ac{CP} framework: the standard confidence given in \cref{def: confidence} (left-hand panel), the conditional signal confidence given in \cref{def: conditional confidence} (middle panel), and the \emph{not-noise} confidence given in \cref{def: not noise} (right-hand panel).
    For each definition, we plot the confidence against the \ac{iFAR} for all triggers (separated by pipeline) in the \ac{MDC}.
    For the standard definition, \cref{def: confidence}, the confidence that the data contains a signal can only be calculated when the single label prediction is for a signal (see \cref{sec: confidence}); we mark these points by a solid line in the left-hand panel.
    Meanwhile, for values of the \ac{iFAR} where the single label prediction is for noise, we use a dashed line.
    We, therefore, see a turnover in the left-most panel, a minimum \ac{iFAR} below which we cannot assign any confidence that the data contains a signal.
    We sort triggers by \ac{iFAR} to produce a continuous line showing the learned mapping.
    In all cases, we truncate the figure at an \ac{iFAR} of $10^4$ years for visualisation purposes: the mapping extends up to the maximum \ac{iFAR} in the data set and monotonically approaches unity in that limit.
    \changed{In the right-hand panel, we add an inset showing the behaviour as each curve approaches unity.}
    }
    \label{fig: MDC11_confidence_FAR}
\end{figure*}

In the \ac{CP} framework, we can use the confidence to assess candidate significance.
As discussed in \cref{sec: confidence}, one can compute either the standard definition of confidence, \cref{def: confidence}, or the conditional confidence, \cref{def: conditional confidence}.
We now consider how these definitions of the confidence can be applied to \ac{CBC} signals using the \ac{MDC} for illustration.

In the left-hand panel of \cref{fig: MDC11_confidence_FAR}, we plot the standard confidence (\cref{def: confidence}) for all triggers in the \ac{MDC} against their \ac{iFAR}.
We find a one-to-one mapping, which is expected since we use the \ac{FAR} as the non-conformity score for the signal label.
The standard confidence that the data contains a signal can only be computed when the single label prediction is for a signal (see \cref{sec: confidence}).
Therefore, we find there is a minimum \ac{iFAR} below which the conditional confidence that the data contains a signal cannot be computed. Instead, we can compute the confidence that the data is noise (since, in this binary case, that is now the single-label prediction).
We illustrate this by adding the noise confidence as a dashed line.

In the middle panel, we go on to show the mapping between the conditional confidence in the signal label, \cref{def: conditional confidence}, against the \ac{iFAR}. Unlike the standard confidence, the conditional confidence can be computed for all values.

For both the standard and conditional confidence, we note that they behave broadly as we expect: the confidence increases monotonically with the \ac{iFAR}.
However, it is notable that the mapping is at odds with the expectation of seasoned analysts in this field: namely, we find that even at a \ac{FAR} of 1 per 1000 years, the confidence of some pipelines is barely above 0.5. 
For comparison, in \cref{fig: FAR-pastro}, at a \ac{FAR} of 1 per 1000 years, all pipelines report a \pastro close to unity.

Moreover, the confidence is pipeline-dependent, with substantial disagreements between pipelines.
This occurs due to our choice of non-conformity score: we use the \ac{FAR}.
The non-conformity score ranks how signal-like the data is compared to the most significant signal in the data: smaller \ac{FAR}s are more signal-like.
As a result, pipelines that have a long tail in the \ac{iFAR} for signals will consequently produce less confidence at the same \ac{iFAR} relative to pipelines with shorter tails.
(It should be remembered, however, at this point that \ac{CP} is distribution-free in the sense that the distributions are never explicit but learned via the calibration data set).
There is nothing inherently wrong here, but we do concur that what is known as confidence in \ac{CP} does not reflect what a gravitational-wave analyst might understand the term to mean.

If we would like the confidence to better reflect our understanding, we can either look at the choice of non-conformity score, or the definition of the confidence. 
An obvious alternative choice for the non-conformity score is \pastro: however, since this is closely related to the \ac{FAR} (c.f. \cref{fig: FAR-pastro}), we encounter similar issues.
Meanwhile, it is worthwhile reflecting on why the seasoned analysts' intuition suggests that a signal with an \ac{iFAR} of 1000 years should confidently be called a signal.
This is because, if the pipeline is well calibrated (which we anticipate to be the case most often), then the \ac{iFAR} intrinsically suggests the data is not consistent with the background.
With this in mind, we define another definition of confidence, the not-noise confidence:
\begin{definition}
\label{def: not noise}
The not-noise confidence is the minimum $1-\alpha$ such that the noise label is not included in $\Gamma^{\alpha}$.
\end{definition}
Applying this definition in the right-hand panel of \cref{fig: MDC11_confidence_FAR}, we recover a mapping much more in line with expectation:
we see a rapid increase in the not-noise confidence, and for values above 1 year, the confidence is close to unity. 
This demonstrates the power of \ac{CP}: it should be remembered that the underlying algorithm is distribution-free, it has learned this intuitive threshold directly from the calibration data.
Moreover, if the underlying algorithm itself was not well calibrated, the confidence still would be (this would manifest as a significant departure from the four calibrated pipelines in the right-hand figure of \cref{fig: MDC11_confidence_FAR}).

The three definitions of confidence presented in \cref{fig: MDC11_confidence_FAR} all offer different ways to assess the confidence we may have in an individual event. However, we believe that further work needs to be done to identify which of these (or perhaps an alternative definition) is best suited to providing a summary of the significance of an individual event. 
Moreover, careful future study will need to be made of how these interact with the choice of non-conformity score.
We also suggest that alternative choices of non-conformity be explored to see if these can better represent our understanding.

%

\begin{figure*}
    \centering
    \includegraphics[width=15cm]{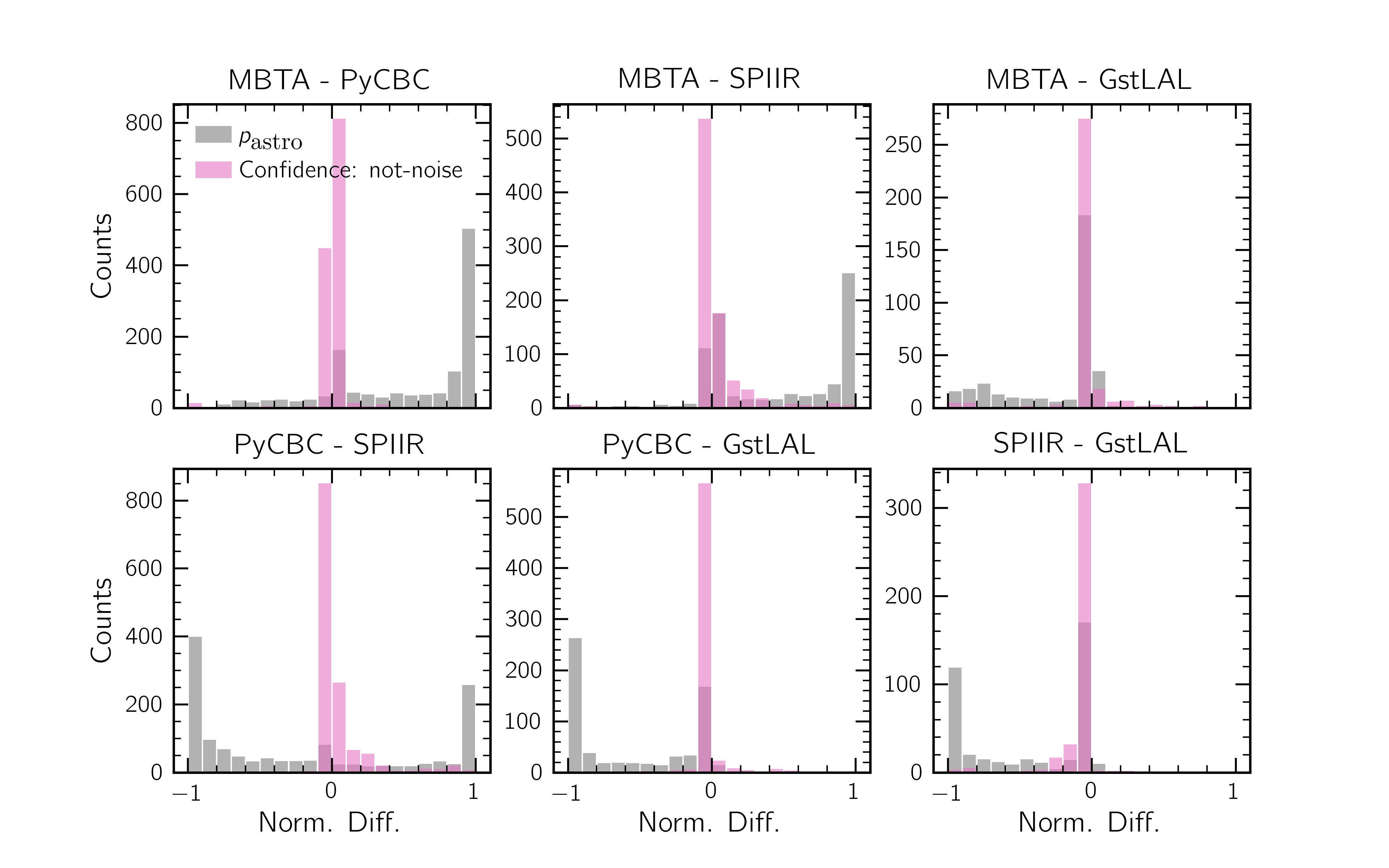}
    \caption{Histogram of the normalised difference (i.e. the difference divided by the sum) in the \emph{not-noise} confidence and \pastro for all pairs of pipelines in the \ac{MDC}.
    Note: we filter to only cases where both pipelines identify the signal (defined as finding a trigger within a $0.1$~s window) and take the closest match in trigger time. We also filter cases where \pastro is not predicted by one or both pipelines.}
    \label{fig: MDC11_pastro_confidence_difference}
\end{figure*}

Finally, if the \ac{CP} calibration has succeeded, we should expect it to regularise pipeline behaviour, i.e. we would expect that the same event found by different pipelines would have a similar confidence.
We would not expect it to give the same confidence to a given event since pipeline performance differs.
To investigate this, in \cref{fig: MDC11_pastro_confidence_difference}, we plot histograms of the normalised difference between the \emph{not-noise} confidence for all pairs of pipelines.
We also show the difference between \pastro for the same pairs.
Notably, while the \pastro difference has a bimodal structure, with frequent cases in which the pipelines completely disagree about a candidate, the confidence difference peaks at zero, demonstrating a spread up to the extremes.
This demonstrates that the confidence measured by \ac{CP} regularises behavior between pipelines by learning from the calibration data set.

\subsection{Conformal Prediction as a generalisation of the traditional framework}
\label{sec: extension}

To conclude our discussion, we finally discuss how the \ac{CP} and traditional \ac{FAR} thresholds are related.
In the traditional framework, to determine if the data contains a signal, we calculate the \ac{FAR} (cf. \cref{eqn: FAR}) and then apply a threshold: $\textrm{FAR}'$.
If the \ac{FAR} is below the threshold, we reject the null hypothesis and determine it is likely a signal.
We can, therefore, formulate this in the language of \ac{CP} by saying that the prediction set of the traditional framework is
\begin{equation}
    \left\{\textrm{``signal''}: \textrm{FAR} < \textrm{FAR}' \right\}\,.
    \label{eqn:pred_trad}
\end{equation}
Formally, this is incorrect as it falls into the ``inverse fallacy'' in that by rejecting the null hypothesis, we assume the data contains a signal. However, in practice, it is very often done.
Meanwhile, in \ac{MCP}, if the ``signal'' non-conformity measure is given by the \ac{FAR} while the ``noise'' non-conformity by the \ac{iFAR}, the prediction set is given by
\begin{equation}
    \left\{\textrm{``signal''}: \textrm{FAR} < \hat{q}_s \right\} \cup
    \left\{\textrm{``noise''}: \textrm{iFAR} < \hat{q}_n \right\} \,,
    \label{eqn:pred_cp}
\end{equation}
where $\hat{q}_s$ and $\hat{q}_n$ are (effectively) the $1-\alpha$ quantile \ac{FAR} and \ac{iFAR} of the calibration data set (cf. \cref{sec: cp}).

Comparing \cref{eqn:pred_trad} and \cref{eqn:pred_cp}, we now see the following three connections between the two methods in the binary classification case where the \ac{FAR} (or equivalently the $p$-value) is used as the non-conformity score.
We use these to explain the differences and advantages of \ac{CP}.

First, in the traditional framework, the threshold for determining if the data contains a signal is chosen by hand. In contrast, in the \ac{CP} framework, the threshold is automatically decided by the algorithm and calibration data set (i.e. $\hat{q}$ is determined by the user choice of $\alpha$).
Of course, if the \ac{FAR} is already well-calibrated, the \ac{CP} framework offers no advantage in this respect. However, if that is not the case, \ac{CP} calibrates the pipeline automatically.

Second, \ac{CP} extends the labelling: while in the traditional framework, one either learns the data is a signal or not, for \ac{CP}, the prediction set can be used to assess significance. I.e. at a fixed choice of $\alpha$, the set may contain both signal and noise: this provides the user with a means to understand the inherent uncertainty, and a choice of definition can be applied to calculate a confidence in a given label.

Finally, we see that in \ac{CP}, one does not fall foul of the inverse fallacy: the signal label arises naturally from the definition of the non-conformity score without assuming it is the negation of the noise label.

Taken together, we therefore argue that \ac{CP} can be viewed as an extension of the traditional statistical framework.

\section{Discussion and applications of conformal prediction}
\label{sec: discussion}
Conformal prediction offers a generalisation of the traditional framework for significance quantification in gravitational-wave astronomy.
In this work, we aim to introduce and explore \ac{CP} in the context of \ac{CBC} searches: we do not seek to demonstrate real application yet and envision this for future work.

We now outline three ways where \ac{CP} may enhance existing efforts.

First, to add the conditional confidence as a calibrated alternative to the \pastro and \ac{FAR} in assessing the significance of single events.
A motivating question we posed in the introduction is how to answer questions such as ``does this data contain an astrophysical signal?''.
The traditional framework answers this by comparing the \ac{FAR} to a threshold or with the astrophysical probability \pastro.
In contrast, \ac{CP} offers the confidence: the key difference between these concepts is that the confidence does not rely on an explicit astrophysical model like the \pastro and is learned from the performance of the pipeline on calibration data.
As shown in \cref{fig: MDC11_pastro_confidence_difference}, this moderates the differences between pipelines, leading to a more stable estimate of the significance.

Second, as a means to automatically set thresholds which guarantee the purity of a catalogue.
With \ac{CP} we can circumvent the problem of determining a threshold on the significance by instead only requiring the user to specify the error rate.
Specifically, given the appropriate tools, a user could set an error rate of 1\% and then take all events where the ``signal'' label is in the prediction set and be assured by \cref{eqn: validity} that at least 99\% of the catalogue are astrophysical signals (within the bounds of the exchangeability assumption).
As shown in \cref{fig: MDC11_error_coverage_bypipeline}, this guarantees the user that the catalogue contains a fixed contamination fraction.

Finally, \ac{CP} offers a framework to develop a post-processing search pipeline combining the outputs from multiple search pipelines.
Specifically, in future work, we will develop a parameterised non-conformity score combining the outputs from multiple pipelines into a single meta-pipeline.
This has the advantage that the between-pipeline behaviour can be regularised using the test and calibration data and we can optimise the score leveraging parameter-space dependent pipeline performance.

For any of these applications to be successful, the critical missing ingredient is a large-scale \ac{MDC}, which accurately captures the actual pipeline performance on realistic data.
The \ac{MDC} used in this work used an unrealistically high event rate and, therefore, is inappropriate for application to astrophysical signals.
Indeed, this underlines the primary limiting factor of \ac{CP}: the assumption of exchangeability between the calibration and test data.
Ensuring this in practice will not be easy.
Unlike many ML use cases, we must simulate the calibration data set for gravitational-wave applications since we do not have a ready training data set.
In the simulation, assumptions must be introduced, e.g., about the waveform models and the rate: assessing and validating these will be critical.
Moreover, using data from past observing runs breaks exchangibility as the detector sensitivity changes dramatically (Moreover, since it changes during an observing run, this is also a concern).
In conformal prediction, such non-exchangeability cases are known as \emph{distribution drift} and can be accounted for by applying weighted conformal procedures \citep{angelopoulos2021gentle}.
Nevertheless, we expect this to be a challenge for any successful application.

We acknowledge that the direction of \ac{CP} is in many respects orthogonal to the overall direction of the field where the \pastro approach has become dominant.
However, we believe that in some cases, end users of the data products do not sufficiently understand the assumptions and caveats of the many \pastro methodologies to interpret them fully.
While $\pastro$ offers a valuable and powerful approach, \ac{CP} offers an alternative in which the end user can, given existing open access to the data and software, calibrate the pipeline themselves, allowing \ac{CP} to learn the uncertainty inherent in the underlying method.
Moreover, we want to emphasise that, for either the \pastro or \ac{FAR} (or equivalently $p$-value approach), if the underlying assumptions are met, \ac{CP} cannot improve on them.
I.e., \ac{CP} does not offer a mechanism to improve the sensitivity of well-calibrated searches. However, it does enable calibration without requiring an understanding of the internal models or making asymptotic assumptions.

Finally, in this work, we have discussed the potential application for \ac{CBC} search.
However, \ac{CP} may also find utility in other areas of the field, such as the low-latency alert products attached to open public alerts, the search for continuous gravitational waves from rapidly rotating neutron stars, or the search for bursts of GWs from unknown sources.

\section{Data Availability Statement}
The source program behind \cref{sec: cpapp} is openly available on Zenodo \citep{Ashton:2023}.

\section{Acknowledgements}
We want to thank Geoffrey Mo for their support in accessing and understanding the \ac{MDC} data products, Will Farr for help with the estimated purity in the \pastro approach, Anarya Ray for support with the \gstlal \pastro values, and Sebastian Khan for comments and corrections to the manuscript.
We also thank Michael Coughlin, Deep Chatterjee, Tito Dal Canton, Reed Essick, Shaon Ghosh, Sushant Sharma-Chaudhary, Max Trevor, and Andrew Toivonen for the development of the MDC results used in this work.
This material is based upon work supported by NSF's LIGO Laboratory which is a major facility fully funded by the National Science Foundation.
Computing support was provided by the Oracle for Research program.

\bibliography{bibliography}
\end{document}